\documentclass[12pt]{article}
\usepackage{latexsym}
\usepackage{amsmath}
\usepackage{amsfonts}
\usepackage{amssymb}
\usepackage[all]{xy}
\usepackage{amscd}
\usepackage{fancybox}
\def\hybrid{\topmargin -20pt    \oddsidemargin 0pt
        \headheight 0pt \headsep 0pt
        \textwidth 6.25in       
        \textheight 9.5in       
        \marginparwidth .875in
        \parskip 5pt plus 1pt   \jot = 1.5ex}

\hybrid

\newcommand{\beq}{\begin{equation}}
\newcommand{\eeq}{\end{equation}}
\newcommand{\bi}{\begin{itemize}}
\newcommand{\ei}{\end{itemize}}
\newcommand{\bea}{\begin{eqnarray}}
\newcommand{\eea}{\end{eqnarray}}
\newcommand{\ba}{\begin{array}}
\newcommand{\ea}{\end{array}}
\newcommand{\bt}{\begin{tabular}}
\newcommand{\et}{\end{tabular}}
\newcommand{\bc}{\begin{center}}
\newcommand{\ec}{\end{center}}

\def\theequation{\arabic{section}.\arabic{equation}}

\newcommand{\vev}[1]{\langle#1\rangle}

\newcommand{\trans}[1]{{\vphantom{#1}}^t{#1}}
\newcommand{\ft}[2]{{\textstyle {\frac{#1}{#2}} }}

\begin{document}

\begin{titlepage}
\begin{center}

\hfill hep-th/0511165\\
\hfill DESY 05-231\\
\hfill ZMP-HH/05-21

\vskip 0.5cm

{\Large \bf On the infinite-dimensional spin-2 symmetries \\ 
in Kaluza-Klein theories  
\\[0.2cm]}

\vskip 1.5cm

{\bf Olaf Hohm} \\

\vskip 20pt

{\em II. Institut f\"ur Theoretische Physik\\ 
Universit\"at Hamburg\\
Luruper Chaussee 149\\
D-22761 Hamburg, Germany\\ \vskip 5pt 
and\\ \vskip 5pt
Zentrum f\"ur Mathematische Physik\\
Universit\"at Hamburg, Bundesstrasse 55\\
D-20146 Hamburg, Germany}

\vskip 15pt

{email: {\tt olaf.hohm@desy.de}} \\

\vskip 0.8cm

\end{center}

\vskip 1cm

\begin{center} {\bf ABSTRACT}\\[3ex]

\begin{minipage}{13cm}
\small
We consider the couplings of an infinite number of spin-2 fields to 
gravity appearing in Kaluza-Klein theories. They are constructed as the 
broken phase of a massless theory possessing an 
infinite-dimensional spin-2 symmetry. 
Focusing on a circle compactification of four-dimensional 
gravity we show that the resulting gravity/spin-2 system 
in $D=3$ has in its unbroken phase 
an interpretation as a Chern-Simons theory of
the Kac-Moody algebra $\widehat{iso(1,2)}$ associated to the Poincar\'e 
group and also fits into the geometrical 
framework of algebra-valued differential geometry developed by Wald.   
Assigning all degrees of freedom to scalar fields,
the matter couplings in the unbroken phase are determined, and it is 
shown that their global symmetry algebra contains the Virasoro algebra 
together with an enhancement of the Ehlers group $SL(2,\mathbb{R})$ 
to its affine extension.
The broken phase is then constructed by gauging a subgroup of the
global symmetries. It is shown that metric, spin-2 fields and 
Kaluza-Klein vectors combine into a Chern-Simons theory for an 
extended algebra, in which the affine Poincar\'e subalgebra 
acquires a central extension.

\end{minipage}
\end{center}

\noindent

\vfill

November 2005

\end{titlepage}

\section{Introduction}
\setcounter{equation}{0}
The AdS/CFT correspondence is still one of the most promising avenues towards
a deeper understanding of the strong coupling mysteries of gauge theories
on the one hand and non-perturbative effects of string or gravity 
theories on the other hand \cite{Maldacena:1997re,Witten:1998qj,
Gubser:1998bc}. 
However, to use the AdS/CFT correspondence 
even at the level where the supergravity approximation is valid, one has 
to compare the full non-linear couplings of in general massive 
Kaluza-Klein states to
operators of corresponding conformal dimension on the CFT side, and 
a restriction to massless modes is therefore unsatisfactory.
Moreover, as the internal manifolds appearing in the background geometries 
of 10-dimensional supergravity, like $AdS_5\times S^5$ or 
$AdS_3\times S^3\times T^4$,
are typically of the same size as the Anti-de Sitter space, the masses 
of the higher Kaluza-Klein states are very small, and it is therefore 
in no sense an effective description to truncate them out.
Nevertheless, usually one indeed inspects only the lowest multiplet, 
simply because of our lack of knowledge concerning the effective actions 
of the higher ones. This paper aims to go a first step towards a 
construction of the full action containing the whole tower of Kaluza-Klein
modes. 

In practice it is a very tedious task even to compute the effective action
for the zero-modes. Thus, in the case of a generic internal manifold 
one would have no chance to compute the action for the
full tower of Kaluza-Klein
modes directly. Instead one therefore may construct the action
directly in the lower-dimensional space-time by identifying the underlying 
symmetries and deriving the possible couplings which are allowed by them. 
This approach has been proven to be very fruitful in 
the construction of gauged supergravities. 
For instance, in case that the number of unbroken supercharges is 
sufficiently large it is usually sufficient to identify
the underlying gauge group, which is given by the isometry group of
the internal manifold, and the spectrum of the various fields under
this gauge group.
More recently this program has been applied to gauged supergravity in
$D=3$ \cite{Nicolai:2003bp,Nicolai:2003ux}.
In this way a tower of massive spin-1 Kaluza-Klein multiplets
for supergravity on $AdS_3\times S^3\times K$ ($K=T^4$ or $K3$)
has been described 
as a unique gauging of three-dimensional supergravity \cite{Nicolai:2003ux},
where the vectors get massive via a three-dimensional version of
the Higgs mechanism. 
What was missing in this description is the full tower of massive spin-2
multiplets (originating from the higher-dimensional metric and its 
superpartners). It is
our aim in this paper to incorporate towers of massive 
spin-2-fields via coupling massless spin-2 fields to gravity and
establishing a novel version of the Higgs mechanism by which they
can become massive.  

However, it is not clear a priori whether such a massless spin-2 theory
actually exists. 
Even though we know that the massive theory appearing via Kaluza-Klein
reduction has to be consistent, the direct construction of 
gravity/spin-2 coupled systems seems to be disabled by the so-called 
interaction problem for higher-spin fields. The latter states that it is 
not clear how to construct an interacting 
theory of higher spin fields (with spin $s>1$), in particular how 
to couple them to gravity (for a recent review 
see \cite{Sorokin:2004ie}.) 
The latter holds for  massive and more 
severely also for massless fields. Let us
illustrate this problem in the 
case of a massless spin-2 field. 
The action for a free massless field of arbitrary spin on a Minkowski 
background\footnote{Even though we are ultimately interested in the AdS case,
we will restrict the analysis to Minkowski backgrounds and discuss 
later on possible generalizations to AdS spaces.}
has been constructed by Pauli and Fierz \cite{Fierz:1939ix} and  
is given in the case of a spin-2 field $h_{\mu\nu}$ by
 \begin{equation}\label{free}
  S=\int d^D x \left[ \frac{1}{2}\partial_{\mu}h_{\nu\rho}
  \partial^{\mu}h^{\nu\rho}-\partial_{\mu}h^{\mu\nu}\partial^{\rho}
  h_{\rho\nu}+\partial_{\mu}h^{\mu\nu}\partial_{\nu}\hat{h}
  -\frac{1}{2}\partial_{\mu}\hat{h}\partial^{\mu}\hat{h}\right], 
 \end{equation}
where $\hat{h}=\eta^{\mu\nu}h_{\mu\nu}$ denotes the trace evaluated 
in the Minkowski metric. This of course coincides with the linearization 
of the Einstein-Hilbert action and  
is moreover invariant under the spin-2 gauge transformations 
$\delta h_{\mu\nu}=\partial_{\mu}\xi_{\nu}+\partial_{\nu}\xi_{\mu}$,
which are the linearized diffeomorphisms. 
To couple this spin-2 field to gravity, naively one would replace all 
partial derivatives in (\ref{free}) by covariant ones. However, this
would violate the invariance under any obvious covariantisation of 
the spin-2 transformation due to the non-commutativity of 
covariant derivatives, $[\nabla_{\mu},\nabla_{\nu}]\sim R_{\mu\nu}$.
This in turn would render the theory 
inconsistent since there is no longer a symmetry which reduces the 
propagating degrees of freedom to the required $\frac{1}{2}D(D-3)$
of spin-2. And indeed it has been proven several times in the literature
that spin-2 fields cannot be coupled consistently 
to gravity (see e.g. \cite{Aragone:1979bm,Boulanger:2000rq,Hindawi:1995an}).
Even though in \cite{Cutler:1986dv,Wald:1986dw} 
Cutler and Wald were able to construct 
interacting multi-graviton theories that are 
consistent in the sense that they admit a spin-2 gauge 
invariance,  later on 
it has been shown \cite{Boulanger:2000rq} that all of these theories 
are unphysical due to the presence of ghost-like excitations. 

The reasoning given for the interaction problem so far 
seems to apply only for massless fields, in which case one has to have
a gauge symmetry, but it is actually more severe in that also a coupling
of massive higher spin fields to gravity is in general not 
consistent. (For the case of massive spin-2 fields a very clear discussion
of the consistency problems can be found in \cite{Hindawi:1995an}, see 
also \cite{Boulware:1973my} and \cite{Buchbinder:1999ar,Buchbinder:2000fy}.)

To make this point more transparent, we remind the reader that
actually the same problem appears for massless spin-3/2 fields $\psi_{\mu}$, 
whose free gauge symmetry $\delta\psi_{\mu}=\partial_{\mu}\epsilon$
is also not consistent with gravity couplings.  
But here it has been partially solved by the invention of local 
supersymmetry. Therefore a number of massless spin-3/2 fields can be
coupled to gravity if they are realized as gauge fields of supersymmetry, 
and moreover, couplings of massive spin-3/2 fields are only 
consistent if they are realized through spontaneously broken supersymmetry.
The massive spin-3/2 fields appearing in 
Kaluza-Klein reductions therefore 
have to be accompanied by spontaneously broken supercharges.
For instance, recently we have constructed the effective supergravity
action for a massive spin-3/2 multiplet
appearing in the KK reduction on $AdS_3\times S^3\times S^3$ 
as a spontaneously broken maximal $N=16$ 
supergravity theory \cite{Hohm:2005ui}, which therefore in total 
has twice as much supersymmetry than is actually preserved by the background.
But the theory admits no symmetric phase, such that
half of the gravitinos get massive through a super-Higgs mechanism.
However, the possible number of spin-$3/2$ fields is limited by the 
maximal amount of $32$ real
supercharges, and it remains the puzzle how an infinite KK tower of 
spin-3/2 fields can be coupled to gravity. 

One of the main aims of the present paper is to analyze in which 
way consistency of the gravity/spin-2 coupling in Kaluza-Klein
theories can be achieved. That means, we will show in which way a 
spin-2 symmetry is realized in the full non-linear theory 
that guarantees consistent couplings in the same sense as 
supersymmetry does for spin-3/2 fields.
More precisely, 
we will construct an unbroken phase with couplings of an infinite 
number of massless spin-2 fields to gravity and possessing a larger 
symmetry structure. This theory will then be deformed via
gauging certain rigid symmetries, which in turn modifies 
the spin-2 symmetry. This is in analogy to the deformation of an ungauged 
into a gauged supergravity, which also modifies
the supersymmetry variations.

In this paper we will focus on a KK reduction to $D=3$. This
is mainly motivated by the topological nature of massless spin-2 fields 
and the fact that the analogous construction
of gauged supergravities in $D=3$ is significantly simplified
due to the fact that in the ungauged theory 
all propagating bosonic degrees of freedom can be assigned 
to scalar fields by use of standard Poincar\'e 
duality \cite{Nicolai:2000sc,Nicolai:2001sv}. 
In fact, starting from a formulation where all vectors have 
been dualized into scalars, the global symmetry group is enhanced, 
e.g. in the maximally supersymmetric case it is the exceptional 
group $E_{8(8)}$. 
Gauging of a subgroup of the global symmetry can in turn be 
performed by just introducing a topological Chern-Simons term 
for the gauge fields (and adding some Yukawa-type couplings).
The analysis of the possible gaugings consistent with supersymmetry
can then be performed in a completely algebraic way.
Fortunately, also gauged supergravities with Yang-Mills type gauging are
covered by these theories, since the on-shell duality
between vectors and scalars actually persists also in the 
gauged phase, as has been shown in \cite{Nicolai:2003bp,deWit:2003ja}.

It will be our aim in this paper to outline a similar procedure for
the construction of the effective action of massive KK
states in $D=3$, 
i.e. we have to show that all degrees of freedom 
can be dualized into scalar 
fields. In an unbroken limit, where the KK masses go to zero,
these will then exhibit an enhanced global symmetry, which severely 
restricts the form of the couplings. 
In addition we will have a local spin-2 symmetry for infinitely many
massless spin-2 fields in much the same way as one has local 
supersymmetry in an ungauged supergravity theory. 
The broken phase of the KK theory will then be constructed by
gauging part of the global symmetry, or in other words by 
switching on a gauge coupling, which will turn out to be 
given by the mass scale $M$ characterizing the inverse radius 
of the internal manifold.
This gauged deformation in turn will induce a mass term for the spin-2 
fields such that a novel Higgs mechanism for the spin-2 fields 
is possible in the same 
way as a super-Higgs effect appears in gauged supergravity.

More specifically, we have to address the following questions: 
 \begin{itemize}
  \item[(i)] 
   How can we identify the unbroken phase, and
   how is the spin-2 symmetry realized in this limit?
   In particular, how does this theory fit into the no-go results discussed
   in the literature before?
  \item[(ii)]  
   Which global symmetry is realized on the scalar fields in this phase? 
  \item[(iii)] 
   Which subgroup of the global symmetries has to be 
   gauged in order to get the full Kaluza-Klein theory?
   Does a formulation exist also for the gauged phase, where all 
   vector and spin-2 fields appear to be topological? 
  \item[(iv)] 
   How does the spin-2 symmetry get modified due to the gauging?
 \end{itemize}

The first step will therefore be to identify the ungauged theory 
with its symmetries, i.e. to answer question (i). 
The required symmetries that appear in Kaluza-Klein reductions have in 
part been analyzed some time ago by Dolan and Duff in \cite{Dolan:1983aa}, 
where they showed that in the simplest case of an $S^1$ compactification
including all massive modes 
a local Virasoro algebra $\hat{v}$ 
corresponding to the diffeomorphisms on $S^1$
as well as the affine extension $\widehat{iso(1,2)}$ 
of the Poincar\'e algebra appear. 
The latter 
describes an infinite-dimensional spin-2 symmetry, 
which in the Kaluza-Klein 
theory is spontaneously broken to the usual diffeomorphism group
such that all higher spin-2
fields get massive via eating vectors and scalars. 
We will construct a consistent infinite-dimensional gauge theory for
massless spin-2 fields as a Chern-Simons theory of $\widehat{iso(1,2)}$ 
and moreover show that the scalar fields span a kind of 
non-linear $\sigma$-model
with target space $\widehat{SL(2,\mathbb{R})}/\widehat{SO(2)}$, exhibiting
an enlarged global symmetry, which answers question (ii).
Answering question (iii) 
we will gauge part of the global symmetry, which in turn 
modifies the spin-2 symmetries. In the formulation, where all 
degrees of freedom appear in scalar fields, the KK vectors 
are topological and we will show that they combine with the spin-2 fields
into a Chern-Simons theory for an extended algebra, which enlightens
the modification of the spin-2 symmetries, thus answering in part 
question (iv).
Schematically, the situation is given below. 

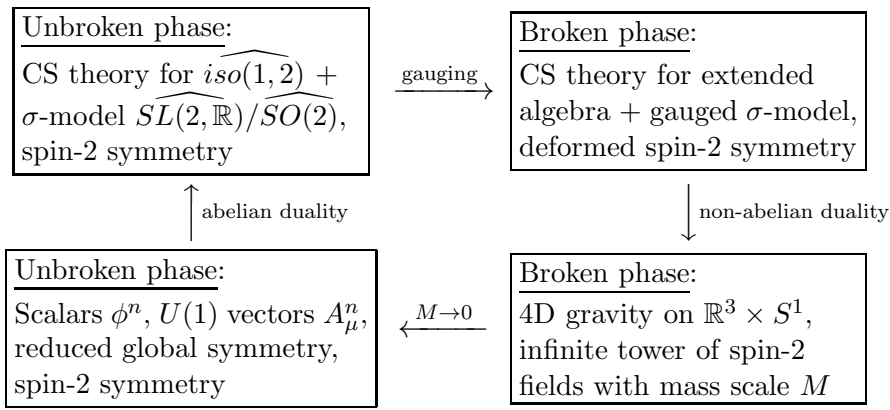
\newenvironment{Boxedminipage}%
{\begin{Sbox}\begin{minipage}}%
{\end{minipage}\end{Sbox}\fbox{\TheSbox}}

\begin{equation*}
\begin{CD}
\begin{Boxedminipage}{4.5cm}
 \small{\underline{Unbroken phase}:  \\ 
  CS theory for $\widehat{iso(1,2)}$ + \\
  $\sigma$-model
  $\widehat{SL(2,\mathbb{R})}/\widehat{SO(2)}$},\\
  spin-2 symmetry
\end{Boxedminipage}
@>\text{gauging } >>
\begin{Boxedminipage}{4.5cm}
 \small{\underline{Broken phase}:  \vspace{0.3em} \\
 CS theory for extended \\algebra
 + gauged $\sigma$-model, \\deformed spin-2 symmetry}
\end{Boxedminipage} \\
@AA{\text{abelian duality}}A
@VV{\text{non-abelian duality}}V\\    
\begin{Boxedminipage}{4.7cm}
 \small{\underline{Unbroken phase}:  \vspace{0.3em} \\
  Scalars $\phi^n$, $U(1)$ vectors $A_{\mu}^n$, \\
  reduced global symmetry, \\
  spin-2 symmetry}
\end{Boxedminipage} 
@<\text{   }M\rightarrow 0\text{   }<< 
\begin{Boxedminipage}{4.5cm} 
 \small{\underline{Broken phase}:  \vspace{0.1em}  \\
  4D gravity on $\mathbb{R}^3 \times S^1$, \\
  infinite tower of spin-2 \\fields with mass scale $M$}
\end{Boxedminipage}
\end{CD}
\end{equation*}

In order to get the full Kaluza-Klein theory (given in 
the lower right corner) we show how to construct the ungauged theory
with its enlarged symmetry group (given in the upper left corner). 
Then we argue that via 
gauging part of the global symmetries we obtain a theory 
(given in the upper right corner) which is on-shell equivalent to 
the original theory.  
However, in the case of a reduction on $S^1$
the effective action can also be computed directly (and in fact has been
computed for reductions to $D=4$, 
see \cite{Cho:1992rq,Cho:1991xk,Cho:1992xv,Aulakh:1985un,Aulakh:1984qx}),
i.e. the unbroken phase could then be 
determined by taking the limit $M\rightarrow 0$ (i.e. obtaining
the theory in the lower left corner) and then performing a standard 
dualization.
Nevertheless this is the simplest case, 
where one can analyze the general aspects 
of the program outlined above, which is crucial 
in order to apply this procedure to
more complicated internal manifolds. 

The paper is organized as follows. After reviewing in sec. 2 the results
of \cite{Dolan:1983aa} applied to the case of an $S^1$ compactification
to $D=3$, the unbroken phase with vanishing KK masses 
will be discussed in sec. 3.
For this we will construct a consistent coupling of infinitely many
massless spin-2 fields to gravity and 
discuss its relation to a geometrical framework
introduced by Wald in \cite{Wald:1986dw}.
Finally, after dualization we analyze the symmetries of the $\sigma$-model, 
which is spanned by the scalar fields. 
In sec. 4 we will turn to the general problem of gauging a 
subgroup of the global symmetries found in sec. 3 and see in particular
how the topological fields combine into an extended Chern-Simons theory,
somehow unifying the internal symmetries with the spin-2 symmetries.
We conclude in sec. 5 with a discussion, including those aspects which
apply also to KK reductions to dimensions other than $D=3$, 
while a generalization to arbitrary internal manifolds will be 
discussed in the appendix.

\section{Kac-Moody symmetries for a Kaluza-Klein theory on 
$\mathbb{R}^3\times S^1$}\setcounter{equation}{0}
It has been shown by Dolan and Duff \cite{Dolan:1983aa} 
that Kaluza-Klein compactification can be analyzed from the
following point of view. The infinite tower of massive modes in the
lower-dimensional Kaluza-Klein spectrum can be viewed as resulting 
from a spontaneous symmetry breaking of an infinite-dimensional 
Kac-Moody-like algebra down to the Poincar\'e group times the 
isometry group of the internal manifold. This infinite dimensional 
symmetry group is a remnant of the higher dimensional diffeomorphism group.   

To be more specific let us review Dolan and Duff's analysis 
applied to the case of a Kaluza-Klein reduction on 
$\mathbb{R}^3\times S^1$. We start from pure Einstein gravity in
$D=4$ and split the vielbein $E_M^A$ in $D=4$ as 
follows\footnote{$M,N,...=0,1,2,5$ denote $D=4$ space-time indices,
$A,B,...$ are flat $D=4$ indices and 
the coordinates are called $x^M=(x^{\mu},\theta /M)$, 
where $M$ is a mass scale characterizing the inverse radius of
the compact dimension. Our metric convention is $(+,-,-)$ for
$D=3$ and similar for $D=4$.}:
 \bea\label{metric}
  E_M^A=\left(\begin{array}{cc} \phi^{-1/2}e_{\mu}^a &
  \phi^{1/2} A_{\mu} \\ 0 & \phi^{1/2} \end{array}\right)\;.
 \eea  
Here we have chosen a triangular gauge and also performed a Weyl rescaling.
The fields are now expanded in spherical harmonics of the 
compact manifold, which for $S^1$ simply reads
 \begin{equation}\label{fieldexp}
  \begin{split}
   e_{\mu}^a(x,\theta)&=\sum_{n=-\infty}^{\infty}
   e_{\mu}^{a(n)}(x)e^{in\theta}\;, \qquad
   A_{\mu}(x,\theta)=\sum_{n=-\infty}^{\infty}A_{\mu}^{n}(x)
   e^{in\theta}\;, \\
   \phi(x,\theta)&=\sum_{n=-\infty}^{\infty}\phi^{n}(x)e^{in\theta},
  \end{split}
 \end{equation}
where we have to impose the reality constraint 
$(\phi^*)^n=\phi^{-n}$ and similarly for the other fields.
Truncating to the zero-modes, the effective Lagrangian is given by
 \begin{equation}\label{zero-mode}
  \mathcal{L}=-eR^{(3)}+\frac{1}{2} 
  eg^{\mu\nu}\phi^{-2}\partial_{\mu}
  \phi\partial_{\nu}\phi-\frac{1}{4}e\phi^2g^{\mu\rho}g^{\nu\sigma}
  F_{\mu\nu}F_{\rho\sigma}\;,
 \end{equation}
where as usual $F_{\mu\nu}$ denotes the $U(1)$ field strength for
$A_{\mu}$. This action is invariant under three-dimensional 
diffeomorphisms and $U(1)$ gauge transformations. 

Let us next analyze how the four-dimensional symmetries are
present in the Kaluza-Klein theory without any truncation. 
For this we notice that the
diffeomorphisms in $D=4$, which are locally generated by a vector 
field $\xi ^M$, are restricted by the topology of the assumed ground 
state $\mathbb{R}^3\times S^1$ to be periodic in $\theta$. 
Therefore we have to expand similarly
 \begin{equation}\label{transexp}
   \xi^{\mu}(x,\theta)=\sum_{n=-\infty}^{\infty}\xi^{\mu (n)}(x)
   e^{in\theta}\;, \qquad
   \xi^5(x,\theta)=\sum_{n=-\infty}^{\infty}\xi^{5 (n)}(x)
   e^{in\theta}\;.
 \end{equation}
The four-dimensional diffeomorphisms act on the vielbein as
 \bea
  \delta_{\xi}E_M^A=\xi^N\partial_NE_M^A + \partial_M\xi^N E_N^A \;,
 \eea
and by applying this formula to (\ref{metric}) we get
 \begin{equation}\label{diff}
  \begin{split}
   \delta_{\xi}\phi &= \xi^{\rho}\partial_{\rho}\phi+\xi^5\partial_5\phi
   +2\phi\partial_5\xi^{\rho}A_{\rho}+2\phi\partial_5\xi^5\;, \\
   \delta_{\xi}A_{\mu}&=\xi^{\rho}\partial_{\rho}A_{\mu}
   +\xi^5\partial_5A_{\mu} + \partial_{\mu}\xi^{\rho}A_{\rho}+\partial_{\mu}
   \xi^5-A_{\mu}\partial_5\xi^{\rho}A_{\rho}-A_{\mu}\partial_5\xi^5\;, \\
   \delta_{\xi}e_{\mu}^a&=\xi^{\rho}\partial_{\rho}e_{\mu}^a+\xi^5\partial_5
   e_{\mu}^a+\partial_{\mu}\xi^{\rho}e_{\rho}^a+\partial_5\xi^5 e_{\mu}^a
   +\partial_5 \xi^{\rho}A_{\rho}e_{\mu}^a\;.
  \end{split}
 \end{equation}
Moreover we have to add a compensating Lorentz transformation with parameter 
$\tau^a_{\hspace{0.3em}5}=-\phi^{-1}\partial_5\xi^{\rho}e_{\rho}^a$
to restore the triangular gauge,
 \bea\label{locomp}
  \delta_{\tau}\phi=0\;, \quad \delta_{\tau}e_{\mu}^a=-A_{\mu}\partial_5
  \xi^{\rho}e_{\rho}^a\;, \quad \delta_{\tau}A_{\mu}=-\phi^{-2}\partial_5
  \xi^{\rho}g_{\rho\mu}\;,
 \eea
where as usual we have written $g_{\mu\nu}=e_{\mu}^ae_{\nu a}$, but now
with $\theta$-dependent vielbein.
Using the mode expansion for the fields in (\ref{fieldexp}) and the
transformation parameter in (\ref{transexp}), one gets an infinite-dimensional
Kaluza-Klein symmetry acting on the fields as\footnote{We will adopt 
the Einstein convention also for double indices 
$m,n=-\infty,...,\infty$, but indicate summations explicitly, if
the considered indices appear more than twice.} 
 \begin{equation}\label{variations}
  \begin{split}
   \delta\phi^n&=\xi_k^{\rho}\partial_{\rho}\phi^{n-k}+iM\sum_k (n+k)\xi_k^5
   \phi^{n-k}+2iM\sum_{k,l}k\xi_k^{\rho}\phi^{n-k-l}A_{\rho}^l\;, \\
   \delta A_{\mu}^n&=
   \partial_{\mu}\xi_n^5+iM\sum_k (n-2k)\xi_k^5A_{\mu}^{n-k}
   +\xi_k^{\rho}\partial_{\rho}A_{\mu}^{n-k}+\partial_{\mu}\xi_k^{\rho}
   A_{\rho}^{n-k}\\ &-iM\sum_{k,l}k\xi_k^{\rho}(\phi^{-2})^{n-k-l}
   g_{\rho\mu}^l-iM\sum_{k,l} k\xi_k^{\rho}A_{\mu}^{n-k-l}A_{\rho}^l \;, \\
   \delta e_{\mu}^{a(n)}&=\xi_k^{\rho}\partial_{\rho}e_{\mu}^{a(n-k)}
   +\partial_{\mu}\xi_k^{\rho}e_{\rho}^{a(n-k)}\\ 
   &+iMn\xi_k^5e_{\mu}^{a(n-k)}
   +iM\sum_{k,l}k\xi_k^{\rho}\left(e_{\mu}^{a(n-k-l)}A_{\rho}^l
   -e_{\rho}^{a(n-k-l)}A_{\mu}^l\right)\;.
  \end{split}
 \end{equation}
Here $(\phi^{-2})^{n}$ is implicitly 
defined by $\phi^{-2}=\sum_{n=-\infty}^{\infty}
(\phi^{-2})^{n}e^{in\theta}$.
 
We see now that the standard Kaluza-Klein vacuum given by the 
vacuum expectation values
 \bea 
  \vev{g_{\mu\nu}}=\eta_{\mu\nu}\;, \qquad \vev{A_{\mu}}=0\;, \qquad 
  \vev{\phi}=1\;,
 \eea
is only invariant under rigid $k=0$ transformations, or in other words 
the infinite-dimensional symmetry is spontaneously broken to the symmetry
of the zero-modes, i.e. to three-dimensional diffeomorphisms and 
a $U(1)$ gauge symmetry.

To explore the group structure which is realized on the whole tower 
of Kaluza-Klein modes including its spontaneously broken part, 
Dolan and Duff proceeded as follows.
Expanding the generators of the $D=3$ Poincar\'e algebra as well as
those for the diffeomorphisms on $S^1$ into Fourier modes, one gets
 \begin{equation}
  P_a^n=e^{in\theta}\partial_a\;, \quad
  J_{ab}^n=e^{in\theta}(x_b\partial_a-x_a\partial_b)\;, \quad
  Q^n=-Me^{in\theta}\partial_{\theta}\;,
 \end{equation}
where we have used flat Minkowski indices.
This implies after introducing $J^a=\ft12\varepsilon^{abc}J_{bc}$
the following symmetry algebra
 \begin{equation}\label{kacmoody}
  \begin{split}
   [P_a^m,P_b^n]&=0\;, \qquad [J_a^m,J_b^n]=\varepsilon_{abc}J^{c(m+n)}\;,
   \qquad [J_a^m,P_b^n]=\varepsilon_{abc}P^{c(m+n)}\;, \\
   [Q^m,Q^n]&=iM(m-n)Q^{m+n}\;, \\
   [Q^m,P_a^n]&=-iMnP_a^{m+n}\;, \qquad [Q^m,J_a^n]=-iMnJ_a^{m+n}\;,
  \end{split}
 \end{equation}
i.e. we get the Kac-Moody algebra associated to the Poincar\'e group 
as well as the Virasoro algebra, 
both without a central extension. 
More precisely, we have a semi-direct product of the Virasoro algebra
$\hat{v}$ with the affine Poincar\'e algebra $\widehat{iso(1,2)}$
in the standard fashion known from the Sugawara 
construction \cite{Goddard:1986bp}.
This algebra should be realized as a local symmetry.
However, this latter statement is a little bit contrived 
since even the diffeomorphisms (i.e. the $k=0$ transformations)
are known not to be realized in general as gauge 
transformations for a certain Lie algebra, e.g. as gauge transformations
for the $P_a$. 
One aim of the present paper is to clarify this question in the 
case of a KK reduction to $D=3$, and we will see how a modification
of (\ref{kacmoody}) appears as a proper gauge symmetry.  

In summary, from this infinite-dimensional symmetry algebra only 
$iso(1,2)\times u(1)$ remains unbroken in the Kaluza-Klein vacuum.
This in turn implies that the fields $A_{\mu}^n$ and $\phi^n$ for 
$n\neq 0$, which correspond to the spontaneously broken generators 
$\xi_n^{\mu}$ and $\xi_n^5$, can be identified
with the Goldstone bosons. They get eaten by the spin-2 
fields $e_{\mu}^{a(n)}$, such that the latter become massive.
A massless spin-2 field carries no local degrees of freedom in $D=3$, 
while a massless vector as well as a real scalar 
each carry one degree of freedom in $D=3$, such that in total 
the massive spin-2 fields each carry two degrees of freedom, 
as expected.

\section{Ungauged phase of the Kaluza-Klein theory}\setcounter{equation}{0}
We have seen in the last section that the full tower of Kaluza-Klein modes
on $\mathbb{R}^3 \times S^1$ carries a representation of an 
infinite-dimensional, but 
spontaneously broken symmetry algebra. 
As indicated in the introduction we are going to construct
the action for this theory by first starting from its unbroken phase, where
all spin-2 fields appear massless and then gauging a certain subgroup of the 
global symmetries.
The unbroken phase is given by the limit, where the 
mass scale $M\rightarrow 0$, or equivalently, where the radius of the 
compact dimension
goes to infinity. In other words, the gauge coupling $g$ will be determined 
by this mass scale, $g=M$.    
In \ref{spin-2} and \ref{geom} we discuss the theory describing the pure
spin-2 sector. Additional matter couplings will be introduced 
in \ref{nonlin}, while possible dualizations are discussed in sec. 3.4.

\subsection{Infinite-dimensional spin-2 theory}\label{spin-2}
To construct the
theory containing infinitely many massless spin-2 fields coupled to gravity,
we remember that according to (\ref{kacmoody}) it should have
an interpretation as a gauge theory of the 
Kac-Moody algebra $\widehat{iso(1,2)}$. 
However, as we already indicated in the discussion at the end of the
previous section, 
in general dimensions 
this is not a helpful statement, because not even pure gravity has an honest
interpretation as a Yang-Mills-like 
gauge theory. 
Fortunately, Witten has shown that in contrast 
gravity in $D=3$ can be viewed 
as a gauge theory of the Poincar\'e algebra $iso(1,2)$, 
namely as a Chern-Simons theory
for this particular non-compact gauge group \cite{Witten:1988hc}. 
Furthermore, the 
symmetries of general relativity, i.e. the diffeomorphisms, are 
on-shell realized as the non-abelian gauge transformations.
We are going to show that correspondingly 
the Chern-Simons theory of the affine $\widehat{iso(1,2)}$ 
describes a consistent coupling of 
infinitely many spin-2 fields to gravity.

To start with, we recall the Chern-Simons theory for a gauge connection 
${\cal A}$, which is given by
 \begin{equation}\label{CS}
  S_{CS}=\int \text{Tr}\big({\cal A}\wedge d{\cal A}+\frac{2}{3}
  {\cal A}\wedge {\cal A}\wedge {\cal A}\big)\;.
 \end{equation}   
Here the trace refers symbolically to an invariant and non-degenerate
quadratic form on the Lie algebra. The invariance of the quadratic form 
$\langle\hspace{0.1em},\rangle$ then implies
that under an arbitrary variation one has
 \begin{equation}\label{var}
  \delta S_{CS}=\int \langle \delta {\cal A}_{\mu},{\cal F}_{\nu\rho}\rangle
  dx^{\mu}\wedge dx^{\nu}\wedge dx^{\rho} \;,
 \end{equation}
where ${\cal F}_{\mu\nu}=\partial_{\mu}{\cal A}_{\nu}
-\partial_{\nu}{\cal A}_{\mu}+[{\cal A}_{\mu},{\cal A}_{\nu}]$ 
denotes the field strength. In particular, 
under a gauge transformation $\delta {\cal A}_{\mu}=D_{\mu}u$,
where $D_{\mu}$ denotes the gauge covariant derivative 
 \begin{equation}
  D_{\mu}u=\partial_{\mu}u+[{\cal A}_{\mu},u]
 \end{equation}
of an infinitesimal transformation parameter $u$,
the action is invariant due to the Bianchi identity. In addition, 
the non-degeneracy of the quadratic form implies the 
equations of motion ${\cal F}_{\mu\nu}=0$.

Therefore, to construct the Chern-Simons theory 
for $\widehat{iso(1,2)}$, we have to find such a quadratic form.
It turns out that
 \bea\label{quadform}
  \langle P_a^m,J_b^n \rangle = \eta_{ab}\delta^{m,-n}\;, \qquad
  \langle P_a^m,P_b^n \rangle = \langle J_a^m,J_b^n \rangle =0
 \eea
defines an invariant form since the bilinear expression
 \bea
  W:=\sum_{n=-\infty}^{\infty}P^{a(n)}J_a^{(-n)}
 \eea
commutes with all gauge group generators. For instance,
 \bea
  [W,P_b^{k}]=\varepsilon_{abc}\sum_{n=-\infty}^{\infty}
  P^{a(n)}P^{c(k-n)}=0
 \eea 
can be seen by performing an index shift
$n\rightarrow n^{\prime}=k-n$, which shows that the sum is symmetric 
in $a$ and $c$.\footnote{Upon truncating the quadratic form to the 
zero-modes, this reduces to an invariant form of the Poincar\'e algebra,
which was the one used in \cite{Witten:1988hc} 
to construct the Chern-Simons action
describing pure gravity in $D=3$.}

We turn now to the calculation of the action, the equations of motion and 
the explicit form of the gauge transformations, which is necessary
to identify the Kaluza-Klein symmetries and fields.
The gauge field takes values in the Kac-Moody algebra,
i.e. it can be written as
 \bea
  {\cal A}_{\mu}=e_{\mu}^{a(n)}P_a^{n}+\omega_{\mu}^{a(n)}J_a^{n}\;.
 \eea
Note, that in the description of ordinary Einstein gravity as 
Chern-Simons theory the gauge field $\omega_{\mu}^a$ is 
interpreted as the spin-connection, 
which like in the Palatini formulation is determined only by the equations 
of motion to be the Levi-Civita connection. Here, instead, we have 
an infinite number of `connections' and their meaning will be 
interpreted later.  

With the invariant quadratic form defined in (\ref{quadform}),
the action reads
 \bea\label{spin2CS}
  S_{CS}=\int d^3 x \:\varepsilon^{\mu\nu\rho}e_{\mu a}^{(n)}
  \big(\partial_{\nu}\omega_{\rho}^{a(-n)}-\partial_{\rho}
  \omega_{\nu}^{a(-n)}+\varepsilon^{abc}\omega_{\nu b}^{(m)}
  \omega_{\rho c}^{(-n-m)}\big)\;.
 \eea
If we define `generalized' curvatures 
 \bea 
  R^{a(n)}=d\omega^{a(n)}+\varepsilon^{abc}\omega^{(m)}_b\wedge
  \omega^{(n-m)}_c\;,
 \eea
the action may be written in a more compact form as
 \bea\label{spin2CS2} 
  S_{CS}=\int e^{(n)}_a\wedge R^{a(-n)}\;.
 \eea
 
The field equations implying vanishing field strength,
${\cal F}_{\mu\nu}=0$, 
read in the given case
 \begin{equation}\label{eom}
  \begin{split}
   \partial_{\mu}e_{\nu}^{a(n)}-\partial_{\nu}e_{\mu}^{a(n)}
   +\varepsilon^{abc}e_{\mu b}^{(n-m)}\omega_{\nu c}^{(m)}
   +\varepsilon^{abc}\omega_{\mu b}^{(n-m)}e_{\nu c}^{(m)}&=0\;, \\
   \partial_{\mu}\omega_{\nu}^{a(n)}-\partial_{\nu}\omega_{\mu}^{a(n)}
   +\varepsilon^{abc}\omega_{\mu b}^{(n-m)}\omega_{\nu c}^{(m)}&=0\;.
  \end{split}
 \end{equation}
Due to the mixing of the infinitely many 'spin connections', the torsion 
defined by $e_{\mu}^{a(0)}$ does 
no longer vanish by the equations of motion. This in turn implies
that it is not transparent which part of the Einstein equation
expresses the curvature and which part the energy-momentum tensor
for the higher spin-2 fields. We will clarify this point later.

Next we evaluate the explicit form of the gauge transformations. 
Introducing the algebra-valued transformation parameter
$u=\rho^{a(n)}P_a^{n}+\tau^{a(n)}J_a^{n}$, for the 
transformations given by $\delta {\cal A}_{\mu}=D_{\mu}u$ one finds
 \begin{equation}\label{kmgauge}
  \begin{split}
   \delta e_{\mu}^{a(n)}&=\partial_{\mu}\rho^{a(n)}+\varepsilon^{abc}
   e_{\mu b}^{(n-m)}\tau_c^{(m)}+\varepsilon^{abc}\omega_{\mu b}^{(n-m)}
   \rho_c^{(m)}\;, \\
   \delta \omega_{\mu}^{a(n)}&=\partial_{\mu}\tau^{a(n)}+\varepsilon^{abc}
   \omega_{\mu b}^{(n-m)}\tau_c^{(m)}\;.
  \end{split}
 \end{equation}
To see that these gauge transformations indeed include the spin-2 
Kaluza-Klein transformations (\ref{variations}) for $M=0$, let us define for
given KK transformation parameterized by $\xi_k^{\mu}$
the gauge parameters
 \bea\label{kkparam}
  \rho^{a(n)}=\xi_k^{\mu}e_{\mu}^{a(n-k)}\;, \qquad
  \tau^{a(n)}=\xi_k^{\mu}\omega_{\mu}^{a(n-k)}\;.
 \eea
Then the gauge transformation (\ref{kmgauge}) takes the form
 \begin{equation}
  \begin{split}
   \delta e_{\mu}^{a(n)}&=\partial_{\mu}\xi_k^{\rho}e_{\rho}^{a(n-k)}
   +\xi_k^{\rho}\partial_{\rho}e_{\mu}^{a(n-k)} \\
   &+\xi_k^{\rho}\left(\partial_{\mu}e_{\rho}^{a(n-k)}
   -\partial_{\rho}e_{\mu}^{a(n-k)}+\varepsilon^{abc}e_{\mu b}^{(n-k-m)}
   \omega_{\rho c}^{(m)}+\varepsilon^{abc}\omega_{\mu b}^{(n-k-m)}
   e_{\rho c}^{(m)}\right)\;,
  \end{split}
 \end{equation}
where we have again performed an index shift.
We see that the first term reproduces the correct KK transformation in 
(\ref{variations}) with $M=0$, 
while the last term vanishes by the equations of
motion (\ref{eom}). On-shell the KK transformations are therefore
realized as gauge transformations. 
That the symmetry is realized only on-shell should not 
come as a surprise because this is already the case for the 
diffeomorphisms \cite{Witten:1988hc}, 
which are now part of the KK-symmetries. 

Thus we have determined a theory which is by construction a consistent 
coupling of infinitely many spin-2 fields.
One might ask the question whether the theory can be consistently 
truncated to a finite number of spin-2 fields, i.e. where only
$e_{\mu}^{a(n)}, n=-N,...,N$ for any finite $N$ remain.
This turns out not to be the case, because setting all generators of
the Kac-Moody algebra with $|n| > N$ to zero
does not result in a consistent Lie algebra since the Jacobi identity
would be violated,\footnote{Similarly it has been shown 
in \cite{Duff:1989ea}, 
that a restriction to a finite number of massive spin-2 fields in Kaluza-Klein
theories is not a consistent truncation.} 
and correspondingly the Chern-Simons theory would be 
inconsistent. 
 
So far we have seen that the action permits
a consistent spin-2 invariance. It remains to be checked that 
it can also be viewed as a 
deformation of a sum of free Pauli-Fierz 
Lagrangians (\ref{free}), in particular that the first-order 
theory constructed here is equivalent to a second order action.
To see this we introduce an expansion 
parameter $\kappa$ and linearize 
the theory by writing
 \bea
  e_{\mu}^{a(0)}=\delta_{\mu}^a+\kappa h_{\mu}^{a(0)}+O(\kappa ^2)\;, 
  \qquad e_{\mu}^{a(\pm 1)}=\kappa h_{\mu}^{a(\pm 1)}+O(\kappa ^2)\;.
 \eea 
We concentrate for 
simplicity reasons only on the case where just 
$e_{\mu}^{a(\pm 1)}$ are present.
Even though, as we have just seen, this is not consistent 
with the gauge symmetry in general, it yields 
correct results up to order $O(\kappa)$ as the corrections
by the full equations are at least of order $O(\kappa^2)$.  
Using the equations of motion for $e_{\mu}^{a(\pm 1)}$ we can now
express $\omega_{\mu}^{a(\pm 1)}$ in terms of them. 
One finds upon expanding up to $O(\kappa)$
 \bea\label{omega1}
   \omega_{\mu}^{a(1)}&=&\kappa\big[ \varepsilon^{\nu ac}(\partial_{\mu}
   h_{\nu c}^{(1)}-\partial_{\nu}h_{\mu c}^{(1)}) 
   +\delta_b^{\nu}(h_{\mu}^{b(1)}\omega_{\nu}^{a(0)}-h_{\mu}^{a(1)}
   \omega_{\nu}^{b(0)}\\ \nonumber
   &&+\omega_{\mu}^{b(0)}h_{\nu}^{a(1)}
   -\omega_{\mu}^{a(0)}h_{\nu}^{b(1)})
   -\frac{1}{4}\varepsilon^{\sigma\rho d}(\partial_{\rho}h_{\sigma d}^{(1)}
   -\partial_{\sigma}h_{\rho d}^{(1)})\delta_{\mu}^a \\ \nonumber
   &&-\frac{1}{4}\delta^{\rho c}\delta^{\sigma d}(h_{\rho d}^{(1)}
   \omega_{\sigma c}^{(0)}-h_{\rho c}^{(1)}
   \omega_{\sigma d}^{(0)}+\omega_{\rho d}^{(0)}h_{\sigma c}^{(1)}
   -\omega_{\rho d}^{(0)}h_{\sigma c}^{(1)})\delta_{\mu}^a\big]
   +O(\kappa^2)\;,
 \eea
and analogously for $\omega_{\mu}^{a(-1)}$. The next step would be 
to insert these relations into the equation for $e_{\mu}^{a(0)}$ and 
solve the resulting expression for the `spin connection' 
$\omega_{\mu}^{a(0)}$. However, due to the fact that $e_{\mu}^{a(\pm 1)}$
as well as $\omega_{\mu}^{a(\pm 1)}$ are of order $O(\kappa)$,
for the approximation linear in $\kappa$ we just get
 \bea 
  0= \varepsilon^{abc}\delta_{\mu b}\omega_{\nu c}^{(0)}+
  \varepsilon^{abc}\delta_{\nu c}\omega_{\mu b}^{(0)}+   
  \kappa\left(\partial_{\mu}h_{\nu}^{a(0)}-\partial_{\nu}h_{\mu}^{a(0)}
  +\varepsilon^{abc}(h_{\mu b}^{(0)}\omega_{\nu c}^{(0)}
  +\omega_{\mu b}^{(0)}h_{\nu c}^{(0)})\right)\;.
 \eea
In the limit $\kappa\rightarrow 0$ this is the relation for vanishing 
torsion in the flat case and is therefore solved by $\omega_{\mu}^{a(0)}=0$.
Up to order $O(\kappa)$ we have the usual relation of vanishing torsion
for the `metric' $h_{\mu}^{a(0)}$ and it can therefore be solved as
in the standard case. But to solve the equation we have to multiply
with the inverse vielbein and therefore up to order $O(\kappa)$
the solution is just the linearized Levi-Civita connection. 
Altogether we have
 \bea
  \omega_{\mu}^{a(0)}=\kappa(\text{lin. Levi-Civita connection})
  +O(\kappa^2)\;.
 \eea
We can now insert this relation into the formulas (\ref{omega1}) for 
$\omega_{\mu}^{a(\pm 1)}$ and get
 \bea
  \omega_{\mu}^{a(1)}=\kappa\big[\varepsilon^{\nu ac}(\partial_{\mu}
  h_{\nu c}^{(1)}-\partial_{\nu}h_{\mu c}^{(1)})-\frac{1}{4}
  \varepsilon^{\sigma\rho d}(\partial_{\rho}h_{\sigma d}^{(1)}
  -\partial_{\sigma}h_{\rho d}^{(1)})\delta_{\mu}^a\big]
  +O(\kappa^2)
 \eea
and analogously for $\omega_{\mu}^{a(-1)}$. 
Finally inserting this expression into the equation of motion for 
$\omega_{\mu}^{a(1)}$ to that order, i.e.
 \bea
  0=\partial_{\mu}\omega_{\nu}^{a(1)}-\partial_{\nu}\omega_{\mu}^{a(1)}\;,
 \eea
and multiplying with $\varepsilon^{\mu\nu\lambda}$ results in 
 \bea\label{freespin2}
  0=\partial_{\mu}\partial^{\lambda}h^{\mu\nu}-\partial^{\nu}
  \partial^{\lambda}\hat{h}+\eta^{\lambda\nu}\square\hat{h}-\eta^{\lambda\nu}
  \partial_{\mu}\partial_{\rho}h^{\mu\rho}+\partial^{\nu}\partial_{\mu}
  h^{\mu\lambda}-\square h^{\nu\lambda}\;.
 \eea 
This coincides exactly with the equation of motion derived from
the original free spin-2 action (\ref{free}).
Here we have defined $h_{\mu\nu}:=\delta_{\mu}^ah_{\nu a}^{(1)}
+\delta_{\nu}^ah_{\mu a}^{(1)}$,
which also results up to $O(\kappa)$ from the general formula
for the higher spin-2 fields in a metric-like representation:
 \bea\label{spin2}
  g_{\mu\nu}^{(n)}:=e_{\mu}^{a(n-m)}e_{\nu a}^{(m)}\;.
 \eea
Clearly, also for
the linearized Einstein equation we get the free spin-2 equation, 
and therefore we can summarize our analysis by saying that the 
theory reduces in the linearization up to order $O(\kappa)$
to a sum of Pauli-Fierz terms.
On the other hand, we know that the full theory has the KK symmetries
(\ref{variations}) which mix fields of different level and 
accordingly the full theory (\ref{spin2CS}) 
has to include non-linear couplings. This in turn implies that 
the higher order terms in $\kappa$ cannot vanish and our theory
is therefore a true deformation of a pure sum of Pauli-Fierz 
terms. In summary, one can solve the equations of motion for
$\omega_{\mu}^{a(n)}$ at least perturbatively, thus giving a second-order
formulation.
However, it would be much more convenient to have a deeper geometrical
understanding for the $\omega_{\mu}^{a(n)}$. Such a geometrical interpretation
indeed exists and is given by Wald's algebra-valued 
differential geometry \cite{Wald:1986dw},
which we are going to discuss in the next section.

\subsection{Geometrical interpretation of the spin-2 symmetry}\label{geom}
Cutler and Wald analyzed in \cite{Cutler:1986dv} the question of 
possible consistent extensions of a free spin-2 gauge invariance 
to a collection of spin-2 fields. 
In much the same way as a non-abelian 
Lie algebra determines 
the gauge symmetry for a collection of spin-1 fields, 
they found that such a spin-2 theory
is organized by an associative and commutative algebra 
$\mathfrak{A}$ (which should not be confused with a Lie 
algebra).
Namely, the additional index which indicates the different 
spin-2 fields is to be interpreted as an algebra index, and therefore
any collection of spin-2 fields can be viewed as a single spin-2 field,
which takes values in a nontrivial algebra.  
Now, an associative and commutative algebra $\mathfrak{A}$ 
can be characterized by its multiplication law, which is with respect 
to a basis given by a tensor $a_{\hspace{0.3em}nm}^{k}$ according to
 \bea
  (v\cdot w)^n = a^n_{\hspace{0.3em}mk}v^mw^k\;,
 \eea
where $v,w \in \mathfrak{A}$.   
That the algebra is commutative and associative is encoded in the 
relations
 \bea\label{asso} 
  a_{\hspace{0.3em}mn}^{k}=a_{\hspace{0.3em}(mn)}^{k}\;, \qquad
  a^k_{\hspace{0.3em}mn}a^n_{\hspace{0.3em}lp}=a^k_{\hspace{0.3em}np}a^n_
  {\hspace{0.3em}ml}\;.
 \eea
With respect to such a given algebra $\mathfrak{A}$, the allowed 
gauge transformations can be written according to \cite{Cutler:1986dv} as 
 \bea\label{conn}
  \delta g_{\mu\nu}^{(n)}=\partial_{(\mu}\xi_{\nu)}^{(n)}
  -2\Gamma_{\mu\nu\hspace{0.6em}l}^{\sigma \hspace{0.3em}n}\xi_{\sigma}^{(l)}
  =:\nabla_{\mu}\xi_{\nu}^{(n)}+\nabla_{\nu}\xi_{\mu}^{(n)}\;,
 \eea
where the generalized Christoffel symbol is defined by
 \bea\label{christ}
  \Gamma_{\mu\nu\hspace{0.6em}l}^{\sigma \hspace{0.3em}n}=
  \frac{1}{2}g^{\sigma\rho\hspace{0.3em}k}_{\hspace{1.5em}l}
  \left(\partial_{\mu}g_{\rho\nu\hspace{0.3em}k}^{\hspace{0.7em}n}
  +\partial_{\nu}g_{\rho\mu\hspace{0.3em}k}^{\hspace{0.6em}n}
  -\partial_{\rho}g_{\mu\nu\hspace{0.3em}k}^{\hspace{0.6em}n}\right)\;,
 \eea
and 
 \bea\label{algmetric}
  g_{\mu\nu\hspace{0.3em}n}^{\hspace{0.6em}k}=
  a_{\hspace{0.3em}nm}^k g_{\mu\nu}^{(m)}\;.
 \eea
We see that $\nabla_{\mu}$ has the formal character of a covariant 
derivative. 

Moreover, it has been shown in \cite{Wald:1986dw} that beyond this
formal resemblance to an ordinary metric-induced connection,
there exists a geometrical interpretation in the
following sense.  
As in pure general relativity, where the symmetry transformations are
given by the diffeomorphisms acting on the fields via the pullback,
the above given transformation rules are the infinitesimal version 
of a diffeomorphism on a generalized manifold. 
This new type of manifold introduced in \cite{Wald:1986dw} generalizes the
notion of an ordinary real manifold to 'algebra-valued' manifolds,
where the algebra $\mathfrak{A}$ replaces the role of $\mathbb{R}$. 
To be more precise, such a manifold is locally
modeled by a $n$-fold cartesian product $\mathfrak{A}^n$ in the
same sense as an ordinary manifold is locally given by $\mathbb{R}^n$.
On these manifolds one can correspondingly define a metric which looks 
from the point of view of the underlying real manifold like an ordinary,
but algebra-valued metric. Now, the diffeomorphisms of these generalized
manifolds act infinitesimally on the metric exactly as written above. 
(For further details see \cite{Wald:1986dw}.)
Moreover, most of the constructions known from Riemannian geometry like 
the curvature tensor have their analogue here.   

To check whether our theory fits into this general framework we
first have to identify the underlying commutative algebra.   
Due to the fact that the theory contains necessarily an infinite 
number of spin-2 fields, the algebra has to be infinite-dimensional, too,
and we will assume that the formalism applies also to this case.
 
We will argue that the algebra is given by the algebra 
of smooth functions on $S^1$, on which we had compactified, 
together with the point-wise multiplication of functions as
the algebra structure.\footnote{That the spin-2 couplings arising
in Kaluza-Klein compactifications might be related to 
Wald's framework in this way 
has first been suggested by Reuter in \cite{Reuter:1988ig},
where he analyzed the reduction of a dimensionally continued Euler form
in $D=6$. Namely, the latter has the exceptional 
property of inducing an infinite
tower of massless spin-2 fields due to the existence of an 
infinite-dimensional symmetry already in the higher-dimensional theory.}
With respect to the complete basis 
$\{e^{in\theta},n=-\infty,...,\infty\}$ of functions on $S^1$, 
the multiplication is given due to elementary Fourier analysis by
 \bea
  (f\cdot g)^{n}=\sum_{m=-\infty}^{\infty} f^{n-m}\cdot g^{m}
  =\sum_{k,m=-\infty}^{\infty}\delta_{k+m,n}f^{k}g^{m}\;,
 \eea  
such that the algebra is characterized by
 \bea\label{alg}
  a_{\hspace{0.3em}km}^{n}=\delta_{k+m,n}\;.
 \eea
This implies that the metric can be written according to (\ref{algmetric}) 
as
 \bea 
  g_{\mu\nu\hspace{0.3em}k}^{\hspace{0.6em}n}=a^n_{\hspace{0.3em}km}
  g_{\mu\nu}^{(m)}=g_{\mu\nu}^{(n-k)}\;.
 \eea
Now it can be easily checked that the KK 
transformations (\ref{variations}) for $M=0$ applied to (\ref{spin2})
can be written as
 \bea 
  \delta g_{\mu\nu}^{(n)}=\nabla_{\mu}\xi_{\nu}^{(n)}
  +\nabla_{\nu}\xi_{\mu}^{(n)}\;,
 \eea
i.e. they have exactly the required form.
Here the connection $\nabla_{\mu}$ is calculated as in (\ref{conn})
with respect to the algebra (\ref{alg}). For this we have 
assumed that indices are raised and lowered according to
 \bea
  \xi_{\mu}^{(n)}=g_{\mu\nu\hspace{0.3em}k}^{\hspace{0.3em}n}\xi^{(k)\nu}\;,
 \eea  
while the inverse metric is defined through the relation
 \bea\label{inverse}
   g^{\mu\rho\hspace{0.3em}n}_{\hspace{1.5em}k}
   g_{\rho\nu\hspace{0.3em}m}^{\hspace{0.6em}k}
   =\delta_{\nu}^{\mu}\delta_m^n\;.
 \eea

With the help of this geometrical interpretation we are now also
able to interpret the existence of an infinite number of 
'spin-connections' $\omega_{\mu}^{a(n)}$.
If we assume that the vielbeins are invertible in the sense of 
(\ref{inverse}), one can solve the 
equations of motion (\ref{eom}) 
for the connections in terms of $e_{\mu}^{a(n)}$, as we have argued in 
sec. \ref{spin-2}.
Then one can define a generalized covariant derivative by postulating 
the vielbein to be covariantly constant,
 \bea\label{covconst}
  \nabla_{\mu}e_{\nu}^{a(n)}=\partial_{\mu}e_{\nu}^{a(n)}-
  \Gamma_{\mu\nu\hspace{0.6em}m}^{\rho \hspace{0.3em}n}e_{\rho}^{a(m)}
  +\omega_{\mu\hspace{0.3em}b}^{a(n-m)}e_{\nu}^{b(m)}=0\;.
 \eea
Since the antisymmetric part $\nabla_{[\mu}e_{\nu]}^{a(n)}$ 
vanishes already by the 
equations of motion (\ref{eom}), this requirement specifies the 
symmetric part of $\nabla_{\mu}e_{\nu}^{a(n)}$. 
In turn, the algebra-valued metric (\ref{spin2}) is covariantly constant
with respect to this symmetric connection,
 \bea
  \nabla_{\mu} g_{\nu\rho}^{(n)}=\partial_{\mu}g_{\nu\rho}^{(n)}
  -\Gamma_{\mu\nu\hspace{0.6em}m}^{\sigma \hspace{0.3em}n}g_{\sigma\rho}^{(m)}
  -\Gamma_{\mu\rho\hspace{0.6em}m}^{\sigma \hspace{0.3em}n}
  g_{\sigma\nu}^{(m)}=0\;.
 \eea
But this is on the other hand also the condition which uniquely fixes the 
Christoffel connection in (\ref{christ}) as
a function of the algebra-valued metric \cite{Wald:1986dw}.
Altogether, the equations of motion for the Chern-Simons action 
(\ref{spin2CS}) together with (\ref{covconst}) determine a 
symmetric connection, which is equivalent to the algebra-valued 
Christoffel connection (\ref{christ}) compatible with the 
metric (\ref{spin2}).

Wald also constructed an algebra-valued generalization of the 
Einstein-Hilbert action, whose relation to the Chern-Simons action 
(\ref{spin2CS}) we are going to discuss now. 
This generalization is (written for three space-times dimensions) 
given by
 \bea
  S^m=\int a^m_{nl}R^n\varepsilon_{\mu\nu\rho}^{\hspace{1.3em}l}
  dx^{\mu}\wedge dx^{\nu}\wedge dx^{\rho}\;,
 \eea
where $R^n$ and $\varepsilon_{\mu\nu\rho}^{\hspace{1.3em}l}$ 
denote the algebra-valued scalar curvature and volume form, 
respectively \cite{Wald:1986dw}.
Applied to the algebra (\ref{alg}) it yields
 \bea\label{algaction}
  S^m=\int R^{m-n}\varepsilon_{\mu\nu\rho}^{\hspace{1.3em}n}
  dx^{\mu}\wedge dx^{\nu}\wedge dx^{\rho}\;.
 \eea
Here the volume form is given by 
$\varepsilon_{\mu\nu\rho}^{\hspace{1.3em}n}=e^n\epsilon_{\mu\nu\rho}$, 
where
 \bea
  e^n=\frac{1}{3!}\epsilon^{\mu\nu\rho}\epsilon_{abc}e_{\mu}^{a(n-m-k)}
  e_{\nu}^{b(m)}e_{\rho}^{c(k)}\;,
 \eea
such that the zero-component of (\ref{algaction})
indeed coincides with the Chern-Simons action 
(\ref{spin2CS2}).
One may wonder about the meaning of the other components of the algebra-valued
action (\ref{algaction}), whose equations of motion 
cannot be neglected for generic algebras.
However, the quadratic form (\ref{quadform}) which has been used
to construct the Chern-Simons action (\ref{spin2CS}), is actually
not unique, but instead there is an infinite series of
quadratic forms,
 \bea\label{quadform2}
  \langle P_a^m,J_b^n \rangle_k = \eta_{ab}\delta^{m,k-n}\;,
 \eea
each of which is invariant and can therefore be used to define 
a Chern-Simons action. These will then be identical
to the corresponding components of the algebra-valued 
action (\ref{algaction}).
But, as all of these actions imply the same equations of motion,
namely ${\cal F}_{\mu\nu}=0$, and are separately invariant under gauge 
transformations, there is no need to consider the full algebra-valued
metric, but instead the zero-component is sufficient. 

Moreover, also matter couplings can be 
described in this framework  in a spin-2-covariant way. 
For instance, an algebra-valued
scalar field $\phi^n$  can be coupled via
 \bea\label{scalar}
  S^m_{\text{scalar}}=\int a^m_{nk}a^n_{lp}\partial_{\lambda}\phi^l
  \partial^{\lambda}\phi^p\varepsilon_{\mu\nu\rho}^{\hspace{1.3em}k}
  dx^{\mu}\wedge dx^{\nu}\wedge dx^{\rho}\;.
 \eea
Furthermore, (\ref{scalar}) is invariant under algebra-diffeomorphisms. 
The latter acts as a Lie derivative
on algebra-valued tensor fields \cite{Wald:1986dw}, 
such that the scalars transform with 
respect to the algebra (\ref{alg}) exactly as required by 
(\ref{variations}) in the phase $M\rightarrow 0$. 
Again, for the algebra (\ref{alg}) considered here, 
the zero-component of (\ref{scalar}) is 
separately invariant and can be written as
 \begin{equation}
  S_{\text{scalar}}=
  \int d^3x \sqrt{g}^{-n}\partial_{\mu}\phi^l\partial^{\mu}\phi^{n-l}
  =\int d^3xd\theta\sqrt{g}\partial_{\mu}\phi\partial^{\mu}\phi\;.
 \end{equation}
Similarly, the Chern-Simons action can be rewritten by retaining a formal
$\theta$-integration and assuming all fields to be $\theta$-dependent.
For explicit computations it is accordingly often more convenient to work with
$\theta$-dependent expressions and therefore we will give subsequent 
formulas in both versions. 

Finally let us briefly discuss the resolution of the aforementioned
no-go theorems for consistent gravity/spin-2 couplings.
In \cite{Boulanger:2000rq} it has been shown that  
Wald's algebra-valued spin-2 theories for arbitrary algebras 
generically contain ghost-like excitations.  
Namely, the algebra has to admit a metric specifying the kinetic 
Pauli-Fierz terms in the free-field limit and moreover has to be symmetric 
in the sense that lowering the upper index in $a_{mn}^k$ by use of 
this metric results in a totally symmetric $a_{mnk}=a_{(mnk)}$.
Now, requiring the absence of ghosts, i.e. assuming the
metric to be positive-definite, restricts the algebra to a  
direct sum of one-dimensional ideals (which means $a_{mn}^k=0$ 
whenever $m\neq n$). The theory reduces in turn to a sum of independent
Einstein-Hilbert terms. For the infinite-dimensional algebra
considered here the metric is given by the $L^2$-norm for 
square-integrable functions (see formula (\ref{L2}) in the appendix),
which is clearly positive-definite. The action may instead be viewed as 
an integral over Einstein-Hilbert terms and is thus in agreement 
with \cite{Boulanger:2000rq}.

\subsection{Non-linear sigma-model and its global symmetries}\label{nonlin}
Apart from the spin-2 sector also the infinite tower of scalar fields
$\phi^n$ will survive in the unbroken limit $M\rightarrow 0$. 
We have seen in the last 
section how spin-2 invariant couplings for scalar fields can be 
constructed. To fix the actual form of these couplings, we have 
to identify also the global symmetries in this limit, and
in order to uncover the maximal global symmetry, we will dualize 
all degrees of freedom into scalars. 

We note from (\ref{variations}) that in the unbroken phase the 
Virasoro algebra $\hat{v}$
parameterized by $\xi_k^5$ reduces to an abelian gauge symmetry, while 
the full $\hat{v}$ will then turn out to be realized only 
as a global symmetry, which is typical for an ungauged limit.
More precisely, we expect an invariance under rigid transformations
of the general form
 \bea\label{rep}
  \delta_{\xi^5}\chi^{n}=i \sum_k (n-(1-\Delta)k)\xi_k^5\chi^{n-k}\;,
 \eea 
where $\xi_k^5$ is now space-time independent. 
One easily checks that these
are representations of $\hat{v}$, which can therefore be labeled by 
their conformal dimension $\Delta$. More precisely, the KK fields 
$e_{\mu}^a$, $A_{\mu}$ and $\phi$ transform as $\Delta=1$, $\Delta=-1$ and  
$\Delta=2$, respectively. 

We start form the zero-mode action (\ref{zero-mode}) and replace it
by the algebra-valued generalization discussed in the last section.
For the Einstein-Hilbert term we have already seen that 
this procedure yields the correct $\widehat{iso(1,2)}$ gauge theory,
and therefore it is sufficient to focus on the scalar kinetic term and 
the Yang-Mills term. The action reads
 \bea\label{matteraction}
  S_{\text{matter}}=\int d^3 xd\theta e\left(-\frac{1}{4}
  \phi^2F^{\mu\nu}F_{\mu\nu}+\frac{1}{2}\phi^{-2}
   g^{\mu\nu}\partial_{\mu}\phi
  \partial_{\nu}\phi\right)\;,
 \eea
where all fields are now $\theta$-dependent or, equivalently, 
algebra-valued.

To dualize the $U(1)$ gauge fields $A_{\mu}^n$ into new 
scalars $\varphi^n$, we define the standard duality relation 
 \bea\label{dual}
  \phi^2F_{\mu\nu}=e\varepsilon_{\mu\nu\rho}g^{\rho\sigma}
  \partial_{\sigma}\varphi\;,
 \eea 
which is not affected by the $\theta$-dependence of all
fields. Thus, the abelian duality between vectors and scalars 
persists also in the algebra-valued case, 
and the degrees of freedom can be assigned to 
$\phi$ and $\varphi$. 
The Lagrangian for the
scalar fields then takes in the unbroken limit $M\rightarrow 0$ 
the form
 \bea
  \mathcal{L}_{\text{scalar}}=\frac{1}{2}eg^{\mu\nu}\phi^{-2}
  (\partial_{\mu}\phi\partial_{\nu}\phi+\partial_{\mu}\varphi
  \partial_{\nu}\varphi)\;,
 \eea
which coincides formally with the zero-mode action after a 
standard dualization, but now with all fields still being 
$\theta$-dependent. 
From (\ref{dual}) one determines the transformation properties of 
the dual scalar $\varphi$ under $\hat{v}$ and finds
 \bea
  \delta_{\xi^5}\varphi=\xi^5\partial_5\varphi+2\varphi\partial_5\xi^5\;,   
 \eea
i.e. it transforms in the same representation as $\phi$ with $\Delta=2$.
(For the computation it is crucial to take into account 
that also $e_{\mu}^a$ transforms under $\hat{v}$.)
Now one easily checks that the action is invariant under global 
Virasoro transformations. 

Moreover, it is well known that the zero-mode scalar fields
span a non-linear $\sigma$-model with coset space $SL(2,\mathbb{R})/SO(2)$ 
as target space, carrying the 'Ehlers group' $SL(2,\mathbb{R})$
as isometry group \cite{Ehlers}. 
If one includes all Kaluza-Klein modes at $M=0$, 
this symmetry is enhanced to an infinite-dimensional algebra, 
which we are going to discuss now. 
Defining the complex scalar field $Z=\varphi+i\phi$, 
the action can be rewritten as
 \bea
  \mathcal{L}_{\text{scalar}}=\frac{1}{2}eg^{\mu\nu}\frac{\partial_{\mu}Z
  \partial_{\nu}\bar{Z}}{(Z-\bar{Z})^2}\;,
 \eea
which is invariant under the $SL(2,\mathbb{R})$ isometries
acting as
 \bea\label{sl2action}
  Z\rightarrow Z^{\prime}=\frac{aZ+b}{cZ+d}\;, 
  \qquad \left(\begin{array}{cc}
  a & b \\ c & d \end{array}\right)\in SL(2,\mathbb{R})\;.
 \eea    
This invariance is not spoiled by the fact that $Z$ is still 
$\theta$-dependent and so the $SL(2,\mathbb{R})$ acts on the full
tower of Kaluza-Klein modes, as can be seen by expanding (\ref{sl2action})
into Fourier modes. But moreover, also the $SL(2,\mathbb{R})$ group 
elements can depend on $\theta$, and therefore an additional 
infinite-dimensional symmetry seems to appear. 

To determine the algebra structure of this infinite-dimensional 
symmetry, let us first introduce a basis for $sl(2,\mathbb{R})$:
 \bea
   h=\left(\begin{array}{cc} 1 & 0 \\ 0 & -1 \end{array}\right), \quad
   e=\left(\begin{array}{cc} 0 & 1 \\ 0 & 0 \end{array}\right), \quad
   f=\left(\begin{array}{cc} 0 & 0 \\ 1 & 0 \end{array}\right).
 \eea
Infinitesimally, with transformation parameter $\alpha=\alpha(\theta)$ 
they act as
 \bea
  \delta_{\alpha}(h)Z=-2\alpha Z\;, \qquad 
  \delta_{\alpha}(e)Z=-\alpha\;, \qquad
  \delta_{\alpha}(f)Z=\alpha Z^2\;,
 \eea
or, expanded in Fourier components, as
 \bea
  \delta_{\alpha^m}(h)Z^n=-2\alpha^m Z^{n-m}\;, \quad
  \delta_{\alpha^m}(e)Z^n=-\delta^{mn}\alpha^m\;, \quad
  \delta_{\alpha^m}(f)Z^n=\alpha^mZ^{n-m-l}Z^l\;.
 \eea
In particular, the real part of $Z$, i.e. the dual scalar $\varphi$, 
transforms as a shift under $e$-transformations, which will later on 
be promoted to local shift symmetries in the gauged theory. 

We can now compute the closure of these symmetry variations
with the Virasoro variations $\delta_{\xi^m}(Q)$.
One finds
 \begin{equation}
  \begin{split}
   [\delta_{\xi^m}(Q),\delta_{\eta^n}(h)]Z^k &=-in\delta_{(\xi\eta)^{m+n}}
   (h)Z^k\;, \\
   [\delta_{\xi^m}(Q),\delta_{\eta^n}(e)]Z^k &=i(-n-2m)
   \delta_{(\xi\eta)^{m+n}}(e)Z^k\;, \\
   [\delta_{\xi^m}(Q),\delta_{\eta^n}(f)]Z^k &=i(-n+2m)
   \delta_{(\xi\eta)^{m+n}}(f)Z^k\;,
  \end{split} 
 \end{equation}
where we have set
 \bea
  (\xi\eta)^{m+n}=\xi^m\eta^n\;.
 \eea
Furthermore, the extended $sl(2,\mathbb{R})$ transformations close 
among themselves according to
 \begin{equation}
  \begin{split}
   [\delta_{\alpha^m}(h),\delta_{\beta^n}(e)]Z^k&=
   2\delta_{(\alpha\beta)^{m+n}}(e)Z^k\;, \\
   [\delta_{\alpha^m}(h),\delta_{\beta^n}(f)]Z^k&=
   -2\delta_{(\alpha\beta)^{m+n}}(f)Z^k\;, \\
   [\delta_{\alpha^m}(e),\delta_{\beta^n}(f)]Z^k&=
   \delta_{(\alpha\beta)^{m+n}}(h)Z^k\;, \\
   [\delta_{\alpha^m}(h),\delta_{\alpha^n}(h)]Z^k&=
   [\delta_{\alpha^m}(e),\delta_{\alpha^n}(e)]Z^k=
   [\delta_{\alpha^m}(f),\delta_{\alpha^n}(f)]Z^k=0\;.
  \end{split}
 \end{equation}
Altogether we can conclude that the following Lie algebra is a global
symmetry of the ungauged theory
 \begin{equation}\label{global}
  \begin{split}
   [Q^m,Q^n]&=i(m-n)Q^{m+n}\;, \quad
   [Q^m,e_n]=i(-n-2m)e_{m+n}\;, \\ [Q^m,h_n]&=-inh_{m+n}\;, 
   \qquad [Q^m,f_n]=i(-n+2m)f_{m+n}\;, \\
   [h_m,e_n]&=2e_{m+n}\;, \qquad [h_m,f_n]=-2f_{n+m}\;, \\ 
   [e_m,f_n]&=h_{m+n}\;, \qquad [e_m,e_n]=[h_n,h_m]=[f_m,f_n]=0\;. 
  \end{split}
 \end{equation} 
We see that the symmetry algebra includes not only the Virasoro algebra 
$\hat{v}$, but also the Kac-Moody algebra 
$\widehat{sl(2,\mathbb{R})}$, 
which transforms under $\hat{v}$. Note, that these
transformation properties are not the standard ones known from the 
Sugawara construction (compare the form of 
$\hat{v}\ltimes\widehat{iso(1,2)}$ in sec. 2, 
see also \cite{Goddard:1986bp}). 
However, this algebra reduces to the standard form upon the 
change of basis given by $\hat{Q}^m=Q^m+mh^m$, such that it clearly defines 
a consistent Lie algebra.

In summary, we can think of the scalar fields $\phi^n$ and $\varphi^n$ 
as parameterizing an
infinite-dimensional $\sigma$-model coset space 
 \bea
  {\cal M}=\frac{\widehat{SL(2,\mathbb{R})}}{\widehat{SO(2)}}\;.
 \eea
Strictly speaking this is not full truth, since the 
metric used to contract indices is actually algebra-valued.
Thus, here we have an algebra-valued generalization of a $\sigma$-model,
which in turn is the reason that it does not only have the 
symmetries $\widehat{sl(2,\mathbb{R})}$, but instead
the whole algebra $\hat{v}\ltimes \widehat{sl(2,\mathbb{R})}$ defined by
(\ref{global}). 

In total, the ungauged phase of the 
effective Kaluza-Klein action without any truncation is therefore  
given by 
 \begin{equation}\label{ungauged}
  \begin{split}
   S=\int d^3 x \:&\big(-\varepsilon^{\mu\nu\rho}e_{\mu a}^{(n)}
   (\partial_{\nu}\omega_{\rho}^{a(-n)}-\partial_{\rho}
   \omega_{\nu}^{a(-n)}+\varepsilon^{abc}\omega_{\nu b}^{(m)}
   \omega_{\rho c}^{(-n-m)}) \\
   &+\frac{1}{2}eg^{\mu\nu}\phi^{-2}
   (\partial_{\mu}\phi\partial_{\nu}\phi+\partial_{\mu}\varphi
   \partial_{\nu}\varphi)\big)\;,
  \end{split}
 \end{equation}
where in the second term 
the algebra multiplication defined in (\ref{scalar}) is implicit, or 
in other words, where all fields are $\theta$-dependent and an integration
over $\theta$ is assumed. 
The action is by construction invariant under spin-2 transformations. 
Moreover, we have already seen that the scalar couplings are also invariant 
under global Virasoro transformations. To see that this is also
the case for the generalized Einstein-Hilbert term,
we have to show that one can determine the 
transformation rule for $\omega_{\mu}^{a(n)}$
such that the action stays invariant.
This is indeed possible, and one finds
 \bea\label{deltaomega}
  \delta_{\xi^5}\omega_{\mu}^{a(n)}=
  i\sum_k (n-k)\xi_k^5\omega_{\mu}^{a(n-k)}\;,
  \qquad \delta_{\xi^5}\omega_{\mu}^a=\xi^5\partial_{\theta}\omega_{\mu}^a\;\;.
 \eea
Equivalently, they can be computed by solving the $\omega_{\mu}^{a(n)}$
in terms of the vielbeins by use of (\ref{eom}) and then applying 
a $\hat{v}$ transformation to this expression. Both results coincide.
Note that instead the full algebra-valued action (\ref{algaction})
transforms non-trivially under $\hat{v}$, namely as 
 \bea
  \delta S^m = im\xi^5_n S^{(m-n)}.
 \eea
However, as we have already seen in sec. 3.2, it is sufficient to include  
only the zero-component in (\ref{ungauged}), which is clearly 
invariant.

\subsection{Dualities and gauging}
So far we have determined the unbroken phase of the KK theory in
a description where all propagating degrees of freedom 
reside in scalar fields.
Before we turn to a gauging of a subgroup of the global symmetries
one may wonder whether it is still possible to assign all degrees of 
freedom to scalars, since the introduction of gauge fields necessarily 
seems to enforce the appearance of local degrees of freedom that  
are instead carried by vectors.   
However, in \cite{Nicolai:2003bp,deWit:2003ja} it has been shown that
in three-dimensional gauged supergravities all Yang-Mills-type gaugings
are on-shell equivalent to Chern-Simons gaugings (with an 
enlarged number of scalar fields), in which case the 
gauge fields are topological and thus all
bosonic degrees of freedom still appear as scalar fields.
We are going to show that this duality also applies to
the present case. 

To begin with, we note that in the gauged theory all partial derivatives 
are replaced by covariant ones. 
For a given field $\chi$ 
transforming in a representation $\Delta$ under $\hat{v}$
the covariant derivative reads 
 \bea
  D_{\mu}\chi^{n}=\partial_{\mu}\chi^{n}-ig\sum_k (n-(1-\Delta)k)A_{\mu}^{k}
  \chi^{n-k}\;,
 \eea
where we have introduced the gauge coupling $g=M$.
Indeed, it transforms by construction covariantly under 
local $\hat{v}$ transformations, $\delta_{\xi}(D_{\mu}\chi^{n})
=ig(n-(1-\Delta)k)\xi_kD_{\mu}\chi^{n-k}$, if we assume as usual that
$A_{\mu}^n$ transforms as a gauge field under the adjoint (i.e. as 
the KK vector in (\ref{variations}) with $\Delta=-1$).
Similarly, the non-abelian $\hat{v}$ field strength is given by 
 \bea
  F_{\mu\nu}^n=\partial_{\mu}A_{\nu}^{n}-\partial_{\nu}A_{\mu}^{n}
  +ig \sum_m (n-2m)A_{\mu}^{n-m}A_{\nu}^{m}\;.
 \eea
These expressions are given for the KK fields in
$\theta$-dependent notation by 
 \begin{equation}\label{covariant}
  \begin{split}
   D_{\mu}\phi&=\partial_{\mu}\phi-gA_{\mu}\partial_{\theta}\phi
   -2g\phi\partial_{\theta}A_{\mu}\;, \\ 
   F_{\mu\nu}&=\partial_{\mu}A_{\nu}-\partial_{\nu}A_{\mu}-gA_{\mu}
   \partial_{\theta}A_{\nu}+gA_{\nu}\partial_{\theta}A_{\mu}\;.
  \end{split}
 \end{equation}

The part of the gauged action containing
scalar fields will be given by the covariantisation of the 
action (\ref{matteraction}) according to (\ref{covariant}).
One easily checks that this transforms into a total $\theta$-derivative 
under local 
$\xi^5_k$-transformations, i.e. defines an invariant action. 
Furthermore, by 
explicit reductions \cite{Cho:1992rq,Cho:1991xk} it has been shown
that exactly these terms appear, as well as an 
explicit $\hat{v}$ gauge invariant mass term 
for the spin-2 fields,  
i.e. the action reads 
 \begin{equation}\label{scalaract}
  \mathcal{L}_{\text{scalar}}=\frac{1}{2} 
  eg^{\mu\nu}\phi^{-2}D_{\mu}
  \phi D_{\nu}\phi-\frac{1}{4}e\phi^2g^{\mu\rho}g^{\nu\sigma}
  F_{\mu\nu}F_{\rho\sigma}+{\cal L}_{\text{mass}}\;.
 \end{equation}

To show that this action is indeed on-shell equivalent to a Chern-Simons
gauged theory we introduce following \cite{Nicolai:2003bp,deWit:2003ja}
new gauge fields for each of the former Yang-Mills fields, 
or in other words, we enhance the gauge symmetry with nilpotent 
shift symmetries (see also \cite{Hohm:2004rc}).
To explain this dualization procedure, let us consider 
the Yang-Mills equation resulting from (\ref{scalaract})
 \bea
  D^{\mu}(\phi^2 F_{\mu\nu})=j_{\nu}\;,
 \eea
where $j_{\nu}$ denotes the current induced by the charged fields. 
It implies integrability of the duality relation
 \bea\label{duality}
  \frac{1}{2} e^{-1}\varepsilon^{\mu\nu\rho}\phi^2F_{\nu\rho}
  =D^{\mu}\varphi+gB^{\mu}
  =:\mathcal{D}^{\mu}\varphi\;,
 \eea
where $\varphi$ will be the scalar field carrying the former degrees of 
freedom of  $A_{\mu}$, and $B_{\mu}$ is the gauge field corresponding to
the enlargement of the gauge group. 
From the previous section we know already the transformation properties
of $\varphi$ under $\hat{v}$, and one may check explicitly 
that it does not change in the gauged phase.
In particular, also the dual vector $B_{\mu}$ will transform 
with $\Delta=2$, in other words it transforms 
under the dual of the adjoint representation of $\hat{v}$,
which will later on turn out to be important.\footnote{For each
representation $\rho$ on a vector space $V$ one has the dual representation
$\rho^{*}$ on the dual space $V^{*}$, which is defined by the requirement
$\left<\rho^*(g)(v^*),\rho(g)(v)\right>=\left<v^*,v\right>$, where 
$\left<,\right>$ denotes the natural pairing between vectors in $V$ and
$V^*$ and $g$ is a group element. This implies 
$\rho^*(g)=\trans{\rho}(g^{-1})$ or at the level of the Lie algebra
$\rho^*(X)=-\trans{\rho}(X)$. Since the adjoint representation
is given by $(t^n)^m_{\hspace{0.3em}k}=f^{nm}_k$, 
the co-adjoint has therefore the matrices $(t_n^*)^{\hspace{0.4em}k}_m
=-(t^n)^k_{\hspace{0.3em}m}=-f^{nk}_m$.}
To define the dual action we add instead of the Yang-Mills term a 
Chern-Simons-like term $B\wedge F$, where $F$ denotes the non-abelian
field strength, and get
 \bea\label{dualscalar}
  \mathcal{L}_{\text{scalar}}=\frac{1}{2}eg^{\mu\nu}\phi^{-2}
  \left(D_{\mu}\phi D_{\nu}\phi+\cal{D}_{\mu}\varphi\cal{D}_{\nu}
   \varphi\right)-\frac{1}{2}g\varepsilon^{\mu\nu\rho}B_{\mu}F_{\nu\rho}
   +{\cal L}_{\text{mass}}\;.
 \eea
Indeed, varying with respect to $B_{\mu}$ one recovers 
the duality relation (\ref{duality}), 
and eliminating the dual scalar $\varphi$ by means
of this relation yields the Yang-Mills type theory (\ref{scalaract}).
Thus we have shown that the degrees of freedom of the $A_{\mu}^n$
can be assigned to new scalars $\varphi^n$, if at the same time 
new topological gauge fields $B_{\mu}^n$ are introduced that promote
the former global shift transformations (i.e. the $e$-transformations 
of $\widehat{sl(2,\mathbb{R})}$) to a local symmetry.

\section{Gauged phase of the Kaluza-Klein theory}\setcounter{equation}{0}
Up to now we have determined the action (\ref{ungauged}) of the ungauged 
theory, which  
is invariant under global $\hat{v}\ltimes \widehat{sl(2,\mathbb{R})}$
transformations as well as local spin-2 transformations.
We argued that in order to get the full Kaluza-Klein action 
one has to gauge the Virasoro algebra together with 
the shift symmetries of (\ref{global}).
In the next section we discuss the 
effect of this gauging on the topological fields. We will see that
they combine into a single Chern-Simons theory, quite  
analogous to gauged supergravities, where truncating to the 
topological fields results in the Chern-Simons theories 
of \cite{Achucarro:1987vz} for $AdS$-supergroups.  
The scalars will be discussed thereafter.

\subsection{Local Virasoro invariance for topological fields}
As usual the gauging proceeds in several steps. First of all, one has to 
replace all partial derivatives by covariant ones. 
Let us start with the generalized Einstein-Hilbert term. 
The covariant derivative for $\omega_{\mu}^{a(n)}$ 
in accordance with (\ref{deltaomega}) reads
 \bea
  D_{\mu}\omega_{\nu}^{a(n)}=\partial_{\mu}\omega_{\nu}^{a(n)}
  -ig\sum_m (n-m)A_{\mu}^m\omega_{\nu}^{a(n-m)}\;,
 \eea
or equivalently
 \bea
  D_{\mu}\omega_{\nu}^{a}=\partial_{\mu}\omega_{\nu}^a
  -A_{\mu}\partial_5 \omega_{\nu}^a\;.
 \eea
The covariantized Einstein-Hilbert action is then invariant under 
local Virasoro transformations. In contrast it will no longer be invariant 
under all spin-2 transformations, but only under three-dimensional
diffeomorphisms, since the explicit $\partial_5$ appearing
in the covariant derivatives will also act on the spin-2 transformation
parameter. Thus, the gauging will deform the spin-2 transformations.
 
Furthermore, we have already seen that 
in order to guarantee that the resulting action will be
equivalent to the original Yang-Mills gauged theory,  
one has to introduce a Chern-Simons term for the 
Kaluza-Klein vectors $A_{\mu}^n$, whose propagating degrees of freedom are
now carried by the dual scalars $\varphi^n$, 
as well as for the dual gauge fields $B_{\mu}^n$.
This implies that we do not have to gauge only the Virasoro algebra
$\hat{v}$, but instead the whole subalgebra of (\ref{global}), which 
is spanned by $(Q^m,e_m)$, while the rigid symmetry given by 
$h_m$ and $f_m$ will be broken explicitly.
Both gauge fields combine
into a gauge field for this larger algebra. 
Moreover, in contrast to 
$\hat{v}$ itself this algebra
carries a non-degenerate invariant 
quadratic form, namely
 \bea\label{virform}
  \langle Q^m, e_n \rangle = \delta^{n,-m}\;,
 \eea  
such that a Chern-Simons action can be defined. The existence 
of this form is due to the fact that $e_m$ transforms actually 
under the co-adjoint action of $\hat{v}$, as we have argued in 3.4.
We will see that the Chern-Simons action with respect to this quadratic
form indeed reproduces the correct $B\wedge F$-term in (\ref{dualscalar}).

It is tempting to ask, whether all topological fields, i.e. the 
gravitational fields together with the gauge fields for the Virasoro and 
shift symmetry, 
can be combined into a Chern-Simons theory for a larger algebra.
The latter would have to combine the affine Poincar\'e 
algebra with the algebra spanned by $(Q^m,e_m)$.
Naively one would think that the semi-direct product 
$\hat{v}\ltimes \widehat{iso(1,2)}$ defined in (\ref{kacmoody}) 
and extended by $e_m$ according to (\ref{global}) is the correct choice.
However, it does not reproduce the right KK symmetry transformations,
and moreover, the algebra seems not to admit a non-degenerate and 
invariant quadratic form.
To see that a Chern-Simons formulation nevertheless exists, we 
observe that varying the total action consisting of the sum of
$\hat{v}$-covariantized Einstein-Hilbert action and $B\wedge F$
with respect to $A_{\mu}$, we get the non-abelian field strength for
$B_{\mu}$ plus terms of the form $e_{\mu}^a\partial_5\omega_{\nu a}$.
Thus, a Chern-Simons interpretation is only possible if the latter
terms are contained in the field strength of $B_{\mu}$, or in
other words, if the algebra also closes according to $[P,J]\sim e$.

Demanding consistency with the Jacobi identities and requiring that
$e_{\mu}^a$ and $\omega_{\mu}^a$ transform under the right representation
of $\hat{v}$, the following Lie algebra is then uniquely fixed up to
a free parameter $\alpha$:
 \begin{equation}\label{bigalgebra}
  \begin{split}
   [P_a^m,J_b^n]&=\varepsilon_{abc}P^{c(m+n)}+i\alpha n\eta_{ab}e_{m+n}\;, 
   \quad [J_a^m,J_b^n]=\varepsilon_{abc}J^{c(m+n)}\;, \\
   [P_a^m,P_b^n]&=0\;, \\
   [Q^m,Q^n]&=ig(m-n)Q^{m+n}\;, \qquad [Q^m,P_a^n]=ig(-m-n)P_a^{m+n}\;, \\
   [Q^m,J_a^n]&=-ignJ_a^{m+n}\;, \qquad [Q^m,e_n]=ig(-n-2m)e_{m+n}\;, \\
   [P_a^m,e_n]&=[J_a^m,e_n]=[e_m,e_n]=0\;.
  \end{split}
 \end{equation}
Here we have rescaled the $Q^m$ with the gauge coupling constant $g$
for later convenience. 

We see that one gets an algebra which looks similar to the one
proposed in \cite{Dolan:1983aa} (see (\ref{kacmoody})), 
except that it does not contain
the semi-direct product of $\hat{v}$ with the affine Poincar\'e algebra,
since the $P_a^n$ and $J_a^n$
transform in different representations of $\hat{v}$.
But in contrast to sec. \ref{nonlin}, where we observed a similar 
phenomenon for the global symmetry algebra, there seems not to exist 
an obvious change of basis which reduces the algebra to the standard form.
Namely, because of the different index structure the $Q^m$ can be 
shifted neither by $P_a^m$ nor $J_a^m$.  
In fact, that the algebra is consistent even in this non-standard form
is possible only because of the nilpotency of translations, 
i.e. $[P_a^m,P_b^n]=0$. 

Furthermore, we observe that the algebra admits a central 
extension $e_m$ of the Poincar\'e algebra
even at the classical level. (Even though, strictly speaking, it is 
only a central extension for the Poincar\'e subalgebra, 
since the $e_m$ do not commute with the $Q^m$.)
Remarkably, it is exactly this modification of the algebra that 
allows the existence of an invariant quadratic form. Namely, the
bilinear expression
 \bea
  W=P^{a(-m)}J_a^{(m)}+\frac{\alpha}{g} Q^m e_{-m}
 \eea
(in particular, $\langle Q^m,e_n\rangle
=\frac{g}{\alpha}\delta^{m,-n}$) 
is invariant under (\ref{bigalgebra}). 
The total Chern-Simons action constructed with respect to this quadratic 
form, with the gauge field written as
 \bea\label{gaugefield}
  {\cal A}_{\mu}=e_{\mu}^{a(n)}P_a^{n}+\omega_{\mu}^{a(n)}J_a^{n}
  +A_{\mu}^n Q^n + B_{\mu}^ne_n\;,
 \eea
is then indeed given by 
 \bea\label{gaugeCS}
  S_{CS}=\int d^3 x d\theta \big(\varepsilon^{\mu\nu\rho}e_{\mu a}
  (D_{\nu}\omega_{\rho}^a-D_{\rho}\omega_{\nu}^a
   +\varepsilon^{abc}\omega_{\nu b}\omega_{\rho c}) 
   +\frac{g}{\alpha}\varepsilon^{\mu\nu\rho}B_{\mu}F_{\nu\rho}\big)\;,
 \eea
i.e. consists of the $\hat{v}$-covariantized Einstein-Hilbert term
and the Chern-Simons action for $A_{\mu}$ and $B_{\mu}$.

Let us briefly comment on the reality constraints on (\ref{bigalgebra}).
Naively one would take (\ref{bigalgebra}) as real Lie algebra, and 
correspondingly the gauge fields in (\ref{gaugefield}) would also be
real. However, the reality condition $(Q^*)^m=Q^m$ (and similarly for 
all other generators) is not consistent,
since taking the complex conjugate of (\ref{bigalgebra}) changes 
relative signs. Instead, only the 
reality constraint $(Q^*)^m=Q^{-m}$ can be consistently imposed. 
This is on the other hand also in accordance with the reality condition 
for the original KK fields in (\ref{fieldexp}), and therefore the fields
in (\ref{gaugefield}) fulfill exactly the correct reality constraint. 
For the Virasoro subalgebra this corresponds to a real section 
where the M\"obius group spanned by $Q^{-1}$, $Q^0$ and $Q^1$ is not
$SL(2,\mathbb{R})$, but instead the compact $SU(2)$.

The equations of motion for the Chern-Simons action in (\ref{gaugeCS}) 
again imply vanishing field strength,
 \bea
  {\cal F}_{\mu\nu}=R_{\mu\nu}^{a(n)}J_a^n+T_{\mu\nu}^{a(n)}P_a^n
  +F_{\mu\nu}^n Q^n + G_{\mu\nu}^n e_n=0\;, 
 \eea 
whose components can in turn be written as 
 \begin{eqnarray}
   R_{\mu\nu}^{a(n)}&=&\partial_{\mu}\omega_{\nu}^{a(n)}-\partial_{\nu}
   \omega_{\mu}^{a(n)}+\varepsilon^{abc}\omega_{\mu b}^{(n-m)}\omega_{\nu c}
   ^{(m)} \nonumber \\
   &+&ig\sum_m (n-m)\omega_{\mu}^{a(n-m)}A_{\nu}^m
   -ig\sum_m mA_{\mu}^{n-m}\omega_{\nu}^{a(m)}\;, \nonumber \\
   T_{\mu\nu}^{a(n)}&=&D_{\mu}e_{\nu}^{a(n)}-D_{\nu}e_{\mu}^{a(n)}
   +\varepsilon^{abc}e_{\mu b}^{(n-m)}\omega_{\nu c}^{(m)}
   +\varepsilon^{abc}\omega_{\mu b}^{(n-m)}e_{\nu c}^{(m)} \;, \nonumber \\
   F_{\mu\nu}^n&=&\partial_{\mu}
   A_{\nu}^{n}-\partial_{\nu}A_{\mu}^{n}
   +ig\sum_m (n-2m)A_{\mu}^{n-m}A_{\nu}^{m}\;, \nonumber \\
   G_{\mu\nu}^n&=&\partial_{\mu}B_{\nu}^n-\partial_{\nu}B_{\mu}^n
   +ig\sum_m(m-2n)A_{\mu}^{n-m}B_{\nu}^m
   +ig\sum_m (n+m)B_{\mu}^{n-m}A_{\nu}^m \nonumber \\ 
   &+& i\alpha \sum_m me_{\mu}^{a(n-m)}\omega_{\nu a}^{(m)}
   - i\alpha\sum_m (n-m)\omega_{\mu}^{a(n-m)} e_{\nu a}^{(m)} \;. 
 \end{eqnarray}
Here $D_{\mu}e_{\nu}^{a(n)}$ denotes the $\hat{v}$-covariant derivative 
on $e_{\mu}^{a(n)}$, which is in $\theta$-notation given by 
 \bea\label{ecov}
  D_{\mu}e_{\nu}^a=\partial_{\mu}e_{\nu}^a-A_{\mu}\partial_5e_{\nu}^a
  -e_{\nu}^a\partial_5 A_{\mu}\;.
 \eea
Moreover, all quantities can be rewritten in $\theta$-dependent notation,
e.g. the non-abelian field strength for $B_{\mu}$ is given by 
 \bea
  \begin{split}
   G_{\mu\nu}&=\partial_{\mu}B_{\nu}-\partial_{\nu}B_{\mu}
   +2(B_{\mu}\partial_5A_{\nu}-B_{\nu}\partial_5A_{\mu})
   -A_{\mu}\partial_5B_{\nu}+A_{\nu}\partial_5B_{\mu} \\
   &+\alpha \left(e_{\mu}^a\partial_{\theta}\omega_{\nu a}
   -e_{\nu}^a\partial_{\theta}\omega_{\mu a}\right) \;.
  \end{split}
 \eea

The gauge transformations for gauge parameter $u=\rho^{a(n)}P_a^n
+\tau^{a(n)}J_a^n+\xi^5_nQ^n+\Lambda^ne_n$ can be written as
 \begin{eqnarray}\label{bigtrans}
   \delta e_{\mu}^{a(n)}&=&\partial_{\mu}\rho^{a(n)}+\varepsilon^{abc}
   e_{\mu b}^{(n-m)}\tau_c^{(m)}+\varepsilon^{abc}\omega_{\mu b}^{(n-m)}
   \rho_c^{(m)}\nonumber \\
   &-&ignA_{\mu}^{n-m}\rho^{a(m)}+igne_{\mu}^{a(n-m)}\xi_m^5 \;,
   \nonumber  \\
   \delta \omega_{\mu}^{a(n)}&=&\partial_{\mu}\tau^{a(n)}+\varepsilon^{abc}
   \omega_{\mu b}^{(n-m)}\tau_c^{(m)} \nonumber \\
   &-&ig\sum_m mA_{\mu}^{n-m}\tau^{a(m)}
   +ig\sum_m (n-m)\omega_{\mu}^{a(n-m)}\xi_m^5 \nonumber \\
   \delta A_{\mu}^{n}&=&\partial_{\mu}\xi^{n}_5
   +ig\sum_m (n-2m)\xi_{m}^5 A_{\mu}^{n-m}\;, \nonumber \\ 
   \delta B_{\mu}^{n}&=&\partial_{\mu}\Lambda^{n}+ig\sum_m (m-2n)\Lambda^{m}
   A_{\mu}^{n-m}+ig\sum_m (n+m)\xi_{m}^5 B_{\mu}^{n-m} \nonumber \\
   &+&i\alpha\sum_m m e_{\mu}^{a(n-m)}\tau_a^{(m)}
   +i\alpha \sum_m (m-n)\omega_{\mu}^{a(n-m)}\rho_a^{(m)} \;.
 \end{eqnarray}

Let us now check, whether the KK symmetries are included in these gauge 
transformations. 
First of all, it reproduces the correct transformation rule for $B_{\mu}$
under $\hat{v}$, as can be seen by rewriting the last equation
of (\ref{bigtrans}) in $\theta$-dependent notation
 \bea
  \delta B_{\mu}=\partial_{\mu}\Lambda-2g\Lambda\partial_{\theta} A_{\mu}
  -gA_{\mu}\partial_{\theta}\Lambda
  +g\xi^5\partial_{\theta} B_{\mu}+2gB_{\mu}\partial_{\theta}\xi^5
  +\alpha e_{\mu}^a\partial_{\theta}\tau_a-\alpha\rho_a
   \partial_{\theta}\omega_{\mu}^a\;.
 \eea
By comparing (\ref{bigtrans}) with (\ref{variations}) we 
also see that the Virasoro gauge transformations
parameterized by $\xi^5$
are correctly reproduced for $e_{\mu}^a$ and $A_{\mu}$. 
To compare with the spin-2 transformations
we define in analogy to (\ref{kkparam}) the transformation parameter
 \bea
  \rho^a=\xi^{\rho}e_{\rho}^a\;, \quad \tau^a=\xi^{\rho}\omega_{\rho}^a
  \;, \quad \xi^5=\xi^{\rho}A_{\rho}\;.
 \eea
Then one finds for the vielbein
 \bea
  \delta_{\xi}e_{\mu}^a=\xi^{\rho}\partial_{\rho}e_{\mu}^a
  +\partial_{\mu}\xi^{\rho}e_{\rho}^a + gA_{\rho}\partial_{\theta}
  \xi^{\rho}e_{\mu}^a
  -gA_{\mu}\partial_{\theta}\xi^{\rho}e_{\rho}^a-\xi^{\rho}T_{\rho\mu}^a\;,
 \eea
which implies that on-shell, i.e. for $T_{\mu\nu}^a=0$, the gauge 
transformations coincide with the KK symmetries in (\ref{diff})
and (\ref{locomp}).
With the same transformation parameter and $\Lambda=\xi^{\rho}B_{\rho}$ 
we find for $A_{\mu}$ and $B_{\mu}$
the following transformation rules (again up to field strength terms)
 \begin{equation}\label{gaugetrans}
  \begin{split}
   \delta_{\xi}A_{\mu} &=\xi^{\rho}\partial_{\rho}A_{\mu}
   +\partial_{\mu}\xi^{\rho}A_{\rho}-gA_{\mu}\partial_{\theta}
   \xi^{\rho}A_{\rho}, \\
   \delta_{\xi}B_{\mu}&=\xi^{\rho}\partial_{\rho}B_{\mu}
   +\partial_{\mu}\xi^{\rho}B_{\rho}-gA_{\mu}\partial_{\theta}
   \xi^{\rho}B_{\rho}
   +2gB_{\mu}\partial_{\theta}\xi^{\rho}A_{\rho}
   +\alpha e_{\mu}^a\partial_{\theta}\xi^{\rho}\omega_{\rho a} \;,
  \end{split}
 \end{equation}
which reproduces for $A_{\mu}$ the same transformation as in 
(\ref{variations}), up to the $\phi$-dependent term (which, of course, 
cannot be contained in a Chern-Simons formulation.)

As in the case of the pure gravity-spin-2 theory, the 
topological phase of the Kaluza-Klein theory is given by a Chern-Simons
theory, and moreover the KK symmetry transformations are on-shell 
equivalent to the non-abelian gauge transformations determined by 
(\ref{bigalgebra}). Even though this equivalence holds only on-shell,
the KK transformations are separately an (off-shell) symmetry, 
since $\delta_{\xi}{\cal A}_{\mu}=\xi^{\rho}{\cal F}_{\rho\mu}$
leaves the Chern-Simons action invariant, as can be easily checked with
(\ref{var}). 

Finally, let us check that spin-2 transformations together with 
the Virasoro transformations build a closed algebra,
as it should be at least on-shell, since they were constructed 
as Yang-Mills gauge transformations.
For the vielbein, e.g., one finds\footnote{As before, we indicate Virasoro
transformations by a subscript $5$ on the transformation parameter.}
 \bea
  [\delta_{\xi},\delta_{\eta^5}]e_{\mu}^a=\delta_{(\eta\xi)}e_{\mu}^a
  -\delta_{(\xi\eta)^5}e_{\mu}^a\;,
 \eea
with the parameter given by 
 \bea\label{param}
  (\eta\xi)^{\rho}=\eta^5\partial_5\xi^{\rho}\;, \qquad
  (\xi\eta)^5=\xi^{\rho}\partial_{\rho}\eta^5\;.
 \eea
The same formula holds for $A_{\mu}$ and $B_{\mu}$. 
But, for $B_{\mu}$ one also has to check the closure of the shift 
symmetries with spin-2 and here one finds
 \bea
  [\delta_{\xi},\delta_{\Lambda}]B_{\mu}=-\delta_{\tilde{\Lambda}}B_{\mu}
  -2\Lambda\partial_5\xi^{\rho}F_{\rho\mu}\;,
 \eea
where 
 \bea\label{shiftclosure}
  \tilde{\Lambda}=\xi^{\rho}\partial_{\rho}\Lambda
  +2\Lambda\partial_5\xi^{\rho}A_{\rho}\;.
 \eea
Therefore the algebra closes only on-shell, i.e. if $F_{\mu\nu}=0$.

\subsection{Virasoro-covariantisation for the scalars}
To summarize the results of the last section, 
we have seen that in the gauged phase the spin-2 
transformations of sec. 3 are no longer a symmetry due to the substitution
of partial derivatives by covariant ones. Therefore
the spin-2 transformations have to be deformed by $g$-dependent 
terms. For the topological fields we have seen that a Chern-Simons
formulation exists, which in turn yields modified spin-2 
transformations, which are consistent by construction.   

Next let us focus on the scalar fields. For them  we 
have already noted the form of the covariant derivative 
in (\ref{covariant}), and
the same formula holds for $\varphi$, but with the difference that
it also has to be covariant with respect to the local shift symmetries
gauged by $B_{\mu}$. The latter act as $\delta_{\Lambda}\varphi=-g\Lambda$, 
i.e. the covariant derivative reads in $\theta$-notation
 \begin{equation}
  \mathcal{D}_{\mu}\varphi=\partial_{\mu}\varphi -A_{\mu}\partial_5\varphi
  -2\varphi\partial_5A_{\mu}+gB_{\mu}\;.
 \end{equation}

Altogether, replacing the partial derivatives in (\ref{ungauged})
by covariant ones and adding the Chern-Simons action constructed in the
last section as well as an explicit mass term which is known 
to appear \cite{Cho:1991xk,Aulakh:1985un}, results in 
 \bea\label{kkaction}
  \begin{split}
   S_{\text{KK}}=\int d^3 x d\theta \big[\varepsilon^{\mu\nu\rho}&
   \big(-e_{\mu}^a
   (D_{\nu}\omega_{\rho a}-D_{\rho}\omega_{\nu a}+\varepsilon_{abc}
   \omega_{\nu}^b\omega_{\rho}^c)
   -\frac{1}{2}gB_{\mu}F_{\nu\rho}\big) \\
   &+\frac{1}{2}eg^{\mu\nu}\phi^{-2}
   (D_{\mu}\phi D_{\nu}\phi+\mathcal{D}_{\mu}\varphi  
   \mathcal{D}_{\nu}\varphi)+{\cal L}_{\text{mass}}\big]\;.
  \end{split}
 \eea
Here we have determined the free parameter of the algebra 
(\ref{bigalgebra}) to be $\alpha=2$ 
in order to get the correct Chern-Simons term for 
$A_{\mu}$ and $B_{\mu}$ discussed in sec. 3.4. 
Namely, there we have already observed 
that varying this action with respect to $B_{\mu}$ one recovers
the duality relation (\ref{duality}). 
In turn, the equations of motion for (\ref{kkaction}) and for the 
Yang-Mills gauged action are equivalent.
This can be seen directly by imposing the gauge $\varphi = 0$
in (\ref{kkaction}) and then integrating out $B_{\mu}$, which
results exactly in the Kaluza-Klein action 
containing the Yang-Mills term in (\ref{scalaract}). 
Moreover, varying with respect to $\omega_{\mu}^a$ still implies 
$T_{\mu\nu}^a=0$. 
This shows that $\omega_{\mu}^a$
can be expressed in terms of $e_{\mu}^a$ as is standard, but with the
exception that all derivatives on $e_{\mu}^a$ are now 
$\hat{v}$-covariant. In the second-order formulation this means that the 
Einstein-Hilbert part looks formally the same as in sec. 3, but
with all Christoffel symbols now containing $\hat{v}$-covariant 
derivatives. This is on the other hand also what one gets by
direct Kaluza-Klein reduction in second-order form 
\cite{Cho:1991xk,Aulakh:1985un}. 
Thus we have shown,
that (\ref{kkaction}) is on-shell equivalent to the Kaluza-Klein
action which results from dimensional reduction. 

In view of the fact that (\ref{kkaction}) is manifestly $\hat{v}$
and shift 
invariant it remains the question how the spin-2 symmetries are realized.
As for the case of the topological fields, 
also the $\sigma$-model action for the scalar fields will no longer 
be invariant under the unmodified spin-2 
transformations for the same reasons. 
To find the deformed transformation rule for the scalars, one way
is to check the closure of the algebra. The unmodified spin-2 
transformations do not build a closed algebra with the local $\hat{v}$ 
transformations. But, if we deform the 
spin-2 transformation to
 \bea
  \delta_{\xi}\phi=\xi^{\rho}\partial_{\rho}\phi
  +2g\phi\partial_{\theta}\xi^{\rho}A_{\rho},
 \eea
the algebra closes according to
 \bea
  [\delta_{\xi},\delta_{\eta^5}]\phi=\delta_{(\eta\xi)}\phi
  -\delta_{(\xi\eta)^5}\phi\;,
 \eea
i.e. exactly like in the case of the topological fields with 
the parameters given in (\ref{param}).
The KK transformations can therefore be entirely reconstructed by 
requiring closure of the algebra. 
The same transformation holds for the dual scalar $\varphi$.

In the presence of matter fields we have to be careful about the 
closure of the algebra also on the gauge fields.
Namely, for the pure Chern-Simons theory shift with spin-2 transformations 
in (\ref{shiftclosure}) close on-shell  (as it
should), but for the theory constructed here the field strength does not
vanish.
Thus one way to get a closing algebra is to extend the 
transformation rule according to
 \bea\label{Bmod}
  \delta^{\prime}B_{\mu}=-2\varphi\partial_{\theta}\xi^{\rho}F_{\mu\rho} \;,
 \eea 
and all transformations close off-shell.

Therefore we see that in the full theory the transformation rules for the 
vectors $A_{\mu}$ and $B_{\mu}$ get extended by scalar field dependent 
terms. That is on the other hand also what we already know from the symmetry 
variations in 
(\ref{variations}) for $A_{\mu}^n$, 
and these terms will be needed in order for 
the full action to be spin-2 invariant. This is in complete analogy
to the construction of gauged supergravities, where the procedure 
of gauging is only consistent with supersymmetry, if additional 
couplings like mass terms are added, while the supersymmetry 
variations are supplemented by scalar-dependent terms.
However, in the present case the invariance of the 
Yang-Mills gauged Kaluza-Klein theory
is guaranteed by construction, which in turn implies that the 
on-shell equivalent dual theory (\ref{kkaction}) is also invariant
(if one assumes transformation rules for $B_{\mu}$, which 
are on-shell given by the variation of the left-hand side of 
(\ref{duality})).
In view of our aim to construct the Kaluza-Klein theories for 
more general backgrounds, it would however 
be important to find a systematic
procedure to determine the scalar-dependent corrections for gaugings of 
arbitrary diffeomorphism Lie algebras.
This we will leave for future work, but here let us 
just show how the scalar-dependent correction in (\ref{Bmod}) 
ensures the invariance under spin-2 for a subsector. 

For this it will be convenient to separate from the spin-2 
transformations those
parts which represent already a symmetry for each term separately.
To do so we remember that to realize the 
spin-2 transformations on the topological fields as gauge 
transformations we had to switch on also the Virasoro transformations
with parameter $\xi^5=\xi^{\rho}A_{\rho}$. Now we will turn the logic
around and apply a spin-2 transformation followed by a Virasoro
transformation with parameter $\xi^5=-\xi^{\rho}A_{\rho}$.
Since Virasoro invariance is manifest, this is a symmetry if and only
if spin-2 is a symmetry. One may easily check that on $e_{\mu}^a$
and $\phi$ (as well as $\varphi$) this transformation is given by 
 \bea\label{gaugeddiff}
  \begin{split}
   \delta_{\xi}\phi&=\xi^{\rho}D_{\rho}\phi\;, \\  
   \delta_{\xi}e_{\mu}^a&=\xi^{\rho}D_{\rho}e_{\mu}^a
   +D_{\mu}\xi^{\rho}e_{\rho}^a\;. 
  \end{split}
 \eea
Here we have used (\ref{ecov})
and also introduced a Virasoro covariant derivative 
for the spin-2 transformation parameter (of which we may
think as transforming as 
$\delta_{\eta^5}\xi^{\mu}=\eta^5 \partial_5\xi^{\mu}$),
 \bea
  D_{\mu}\xi^{\rho}=\partial_{\mu}\xi^{\rho}-A_{\mu}\partial_5\xi^{\rho} \;.
 \eea
We see that we get transformation rules which look formally like
a diffeomorphism symmetry, except that all appearing derivatives 
are $\hat{v}$-covariant. In the following we will refer to these
transformations as `gauged diffeomorphisms'.  
In contrast, the gauge fields $A_{\mu}$ and $B_{\mu}$ transform as
 \bea\label{Adiff}
  \begin{split}
   \delta_{\xi}A_{\mu}&=\xi^{\rho}F_{\rho\mu}\, \\
   \delta_{\xi}B_{\mu}&=
   \xi^{\rho}D_{\rho}B_{\mu}+D_{\mu}\xi^{\rho}B_{\rho}
   +\alpha e_{\mu}^a\partial_{\theta} \xi^{\rho}\omega_{\rho a}\;. 
  \end{split} 
 \eea

It remains the question whether actions can be constructed that are 
manifestly invariant under these transformations.
To analyze this let us start with an action constructed from a scalar 
Lagrangian given by
 \bea
  S=\int d^3 x d\theta \hspace{0.2em}e\hspace{0.2em} \mathcal{L}\;,
 \eea
and moreover being invariant under local Virasoro transformations.
Put differently, this means that the Lagrangian varies as 
$\delta_{\xi^5}{\cal L}=\xi^5\partial_5{\cal L}-2{\cal L}\partial_5\xi^5$
under $\hat{v}$ (because then it transforms together with the 
vielbein determinant, whose symmetry variation reads 
$\delta_{\xi^5}e=\xi^5\partial_5 e +3e\partial_5\xi^5$, 
into a total $\theta$-derivative). 
By use of the $\hat{v}$-covariant derivative 
given by 
 \bea
  D_{\mu}\mathcal{L}=\partial_{\mu}{\cal L}-A_{\mu}\partial_5{\cal L}
  +2{\cal L}\partial_5 A_{\mu}\;,
 \eea
we can then evaluate the variation of the action under gauged 
diffeomorphisms and find
 \bea
  \begin{split}
   \delta_{\xi} S &=\int d^3 x d\theta \left[(\xi^{\rho}D_{\rho}e
   +eD_{\rho}\xi^{\rho})\mathcal{L}+e\xi^{\rho}D_{\rho}\mathcal{L}\right]
   = \int d^3 x d\theta D_{\rho}(e\xi^{\rho}\mathcal{L})\\
   &=\int d^3 x d\theta\left[\partial_{\rho}(e\xi^{\rho}\mathcal{L})
   -\partial_5(e\xi^{\rho}A_{\rho}\mathcal{L})\right]=0\;.
  \end{split}
 \eea
Thus, if one constructs an action from a Lagrangian that 
transforms as a scalar under gauged diffeomorphisms, then the action
is invariant under these gauged diffeomorphisms if and only if it
is also invariant under local Virasoro transformations. 
The latter requirement is satisfied in our theory by construction.
Thus it remains to be checked whether the Lagrangian transforms as a
scalar. However, using (\ref{gaugeddiff}), (\ref{Adiff}) 
and $[D_{\mu},D_{\nu}]\phi=
-2\phi\partial_5 F_{\mu\nu}-\partial_5\phi F_{\mu\nu}$, one proves
that the covariant derivative $D_{\mu}\phi$  
transforms under gauged diffeomorphisms as 
 \bea\label{covdiff}
  \delta_{\xi}(D_{\mu}\phi)=D_{\mu}\xi^{\rho}D_{\rho}\phi
  +\xi^{\rho}D_{\rho}D_{\mu}\phi-2\phi\partial_5\xi^{\rho}F_{\rho\mu}\;,
 \eea
i.e. it does not transform like a one-form, but requires 
an additional piece proportional to $F_{\mu\nu}$, 
which again shows that corrections have
to be added to the transformation rules.  

Let us consider the subsector of the theory where we rescale
 \bea
  e_{\mu}^a\rightarrow \kappa e_{\mu}^a\;, \qquad
  \varphi \rightarrow \kappa ^{-1/2}\varphi\;,
 \eea
and then take the limit $\kappa \rightarrow 0$. 
The action then reads
 \bea\label{subsec}
  S_{\kappa\rightarrow 0}=\frac{1}{2}\int d^3 xd\theta\left(-g
  \varepsilon^{\mu\nu\rho}B_{\mu}F_{\nu\rho}+eg^{\mu\nu}\phi^{-2}
  {\cal D}_{\mu}\varphi{\cal D}_{\nu}\varphi\right)\;.
 \eea
However, in spite of the fact that the term $\sim e_{\mu}^a
\partial_{\theta} \xi^{\rho}\omega_{\rho a}$ in (\ref{Adiff}) 
disappears in this limit 
and with the additional contribution (\ref{Bmod}) in 
the $B_{\mu}$ variation, the extra term in (\ref{covdiff}) is 
canceled, and the kinetic term for $\varphi$ is therefore separately
invariant. The Chern-Simons term on the other hand transforms according
to (\ref{var}) as
  \bea
  \delta_{\xi}S_{\kappa\rightarrow 0}
  =-g\int\varepsilon^{\mu\nu\rho}F_{\sigma\mu}F_{\nu\rho}\varphi
  \partial_5\xi^{\sigma}=0\;,
 \eea
where we have used that a totally antisymmetric object in four indices 
vanishes in $D=3$.
Thus we have shown the scalar field modification in (\ref{Bmod}) 
is sufficient in order to restore the spin-2 invariance of this
subsector of the theory.

\section{Conclusions and Discussion}
In this paper we have examined the spin-2 symmetries appearing in 
Kaluza-Klein theories. Via analyzing an $S^1$ compactification to $D=3$
we have shown how a spin-2 symmetry is realized in the unbroken phase
by constructing the Chern-Simons theory of the affine Poincar\'e algebra.
This spin-2 symmetry fits also into a geometrical framework introduced 
by Wald. The latter associates to any given commutative algebra a 
consistent multi-graviton theory, and in the present case the 
algebra is given by the algebra of smooth functions on $S^1$.  
Even though the Chern-Simons description is special to $D=3$,
the latter construction is possible in any dimension. 
Moreover, by constructing the analogous extension of the Poincar\'e
algebra for arbitrary internal manifolds
we prove in the appendix that also in these cases  
the Chern-Simons theories are equivalent to the algebra-valued gravity
theories. Thus we have confirmed and refined the proposal made 
in \cite{Reuter:1988ig}.

In addition, we discussed matter-couplings in the unbroken 
phase in a formulation, 
where all degrees of freedom reside in scalar fields, 
and showed that they admit an enhanced infinite-dimensional
rigid symmetry, which contains the Virasoro algebra and an affine 
extension of the Ehlers group. 

Upon gauging part of the global symmetries we constructed the 
broken phase, which adds a Chern-Simons term for the KK vectors
and an explicit mass term for the spin-2 fields. 
The latter theory was shown to be on-shell equivalent to the Yang-Mills
gauged action which one gets by direct KK reduction. 
Moreover, the KK vectors combined with the metric and the higher spin-2 fields
into a Chern-Simons theory for an extended algebra, which 
features also a central extension for the affine Poincar\'e subalgebra.
The non-abelian gauge transformations for this theory were shown
to be on-shell equivalent to the KK transformation in (\ref{variations}),
up to scalar-dependent terms.
This proves that truncating only to the topological fields
and disregarding scalar field contributions results in a consistent 
theory. The latter is in analogy to gauged supergravities in $D=3$,
where a truncation to the supergravity multiplet and the compact 
Chern-Simons gauge fields results in the pure Chern-Simons theories
for $AdS$-supergroups constructed in \cite{Achucarro:1987vz}.
Similarly, in \cite{Blencowe:1988gj} couplings of an infinite 
tower of higher-spin
fields to gravity have been constructed as Chern-Simons theories for 
higher-spin algebras, and the present work contains
the analogue for an infinite tower of spin-2 fields.  

Based on the algebra (\ref{bigalgebra}) we were able to analyze 
how the gauging deforms the spin-2 symmetries
for the topological fields.
But, we can no longer expect this algebra to be realized as a
proper gauge symmetry also on the matter fields, since here even ordinary
diffeomorphisms have no direct 
interpretation as Poincar\'e gauge transformations.
We have seen that also scalar-dependent 
modifications of the spin-2 transformations are required 
and have shown explicitly how the spin-2 invariance of a 
subsector (\ref{subsec}) of the theory is restored.
However, we did not analyze in detail which scalar-dependent 
deformations guarantee the invariance of the full theory containing
a spin-2 mass term (whose form 
can be found in \cite{Cho:1992rq}),
but just observed that the existence of such a deformation is ensured 
by the on-shell equivalence to the original KK theory. 
We will leave a more exhaustive analysis of these deformations, 
i.e. a complete answer to question (iv) raised in the introduction,
for future work. The latter will also be necessary in order to 
consider more general internal manifolds \cite{Aulakh:1985my}.   
 
Finally let us comment on possible generalizations. 
Since our original motivation was given by the AdS/CFT 
correspondence, we would like to analyze  
KK theories on $AdS_3$. For this, naively one would first replace 
the Poincar\'e algebra in (\ref{bigalgebra}) by the $AdS_3$-isometry
group $SO(2,2)$.
But for non-vanishing cosmological constant $\lambda$ one would get
the additional commutator $[P_a,P_b]\sim \lambda J_{ab}$, which 
in turn would render the algebra (\ref{bigalgebra}) 
with its non-standard product 
between $\hat{v}$ and the Kac-Moody algebra inconsistent,
as its consistency relied heavily on the 
nilpotency of the $P_a^m$. 
That this algebra would be inconsistent should not be a surprise,
because it would correspond to a compactification on $AdS_3\times S^1$,
which is not a solution of four-dimensional AdS-gravity. 
Accordingly, the naive $S^1$-reduction of AdS-gravity in $D=4$ 
would result in a cosmological constant term in $D=3$ containing
an explicit dilaton factor $\phi$, which therefore could 
clearly not be covered by a Chern-Simons description. 
On the other hand, for compactifications on, say, $AdS_3\times S^3$,
which are solutions of the higher-dimensional gravity theories,
one would expect the appearance of consistent algebras like 
(\ref{bigalgebra}), containing an affine extension of $SO(2,2)$.

Moreover, it would be interesting to extend the analysis to the 
compactification of supergravity theories \cite{Dolan:1984fm}. 
This would require an 
extension of (\ref{bigalgebra}) to a super-Kac-Moody algebra, 
whose Chern-Simons theory would 
then describe a supergravity 
with an infinite number of gravitinos, i.e. possessing an 
infinite number of supercharges ($N=\infty$).
A resolution of the apparent conflict with the no-go theorem
restricting the number
of real supercharges to be less or equal to $32$, is given by
the observation that upon coupling to propagating matter 
these theories need not to admit a phase where the whole 
supersymmetry is unbroken 
(see also the discussion in \cite{Hohm:2005ui}).

\subsection*{Acknowledgments}
I am greatly indebted to Henning Samtleben for stimulating discussions
at every stage of this project and for carefully reading the manuscript.
Moreover, I would like to thank Christian Becker, Jens Fjelstad 
and Mattias Wohlfarth for useful conversations.  
This work is supported by the EU contracts MRTN-CT-2004-503369 and
MRTN-CT-2004-512194, the DFG grant SA 1336/1-1 and 
DAAD -- The German Academic Exchange Service.

\begin{appendix}
\renewcommand{\theequation}{\Alph{section}.\arabic{equation}}

\section{Appendix: Spin-2 theory for arbitrary 
internal manifold}\setcounter{equation}{0}\label{A}  
In the main text we have shown that the Chern-Simons gauge theory
of the affine Poincar\'e algebra describes a consistent 
gravity-spin-2 coupling. This is on the other hand also equivalent to
Wald's algebra-valued generalization of the Einstein Hilbert action,
where the algebra is given by the algebra of smooth functions 
on $S^1$. 
We are going to show that this picture generalizes to the case of
an arbitrary internal manifold. 
 
Let $M$ be an arbitrary compact Riemannian manifold and 
$\{e_m\}$ a complete set of spherical harmonics (where generically 
$m$ now denotes a multi-index), which we also take as a basis for 
the algebra of smooth functions on $M$. 
The infinite-dimensional extension
of the Poincar\'e algebra is no longer given by a Kac-Moody algebra, but 
instead is spanned by generators $P_a^n=P_a\otimes e_n$ and
$J_a^n=J_a \otimes e_n$, which satisfy the Lie algebra (compare the 
algebra in \cite{Boulanger:2002bt})
 \bea\label{genalg}
  [P_a^m,P_b^n]=0\;, \quad 
  [J_a^m,J_b^n]=\varepsilon_{abc}J^{c}\otimes (e_m \cdot e_n)\;,
   \quad [J_a^m,P_b^n]=\varepsilon_{abc}P^{c}\otimes (e_m \cdot e_n)\;.
 \eea  
Here $\cdot$ denotes ordinary multiplication of functions.
Note, that this algebra reduces for the case $M=S^1$ to the
Kac-Moody algebra $\widehat{iso(1,2)}$ in (\ref{kacmoody}).
There exists also an inner product on the space of functions, which
is given by
 \bea\label{L2}
  (e_m,e_n)=\int_M \text{dvol}_M e_m e_n\;,
 \eea
such that a non-degenerate quadratic form on (\ref{genalg}) 
exists:
 \bea
  \langle P_a^m,J_b^n \rangle =\eta_{ab} (e_m,e_n)\;.
 \eea
In complete analogy to sec. 3 a Chern-Simons theory can then 
be defined, whose equations of motion read 
 \begin{equation}
  \begin{split}
   \partial_{\mu}e_{\nu}^{a(n)}-\partial_{\nu}e_{\mu}^{a(n)}
   +a^n_{mk}\varepsilon^{abc}\left(e_{\mu b}^{(m)}\omega_{\nu c}^{(k)}
   +\omega_{\mu b}^{(m)}e_{\nu c}^{(k)}\right)&=0\;, \\
   \partial_{\mu}\omega_{\nu}^{a(n)}-\partial_{\nu}\omega_{\mu}^{a(n)}
   +a^n_{mk}\varepsilon^{abc}\omega_{\mu b}^{(m)}\omega_{\nu c}^{(k)}&=0\;, 
  \end{split}
 \end{equation}
while they are invariant under
 \bea
   \delta e_{\mu}^{a(n)}=\partial_{\mu}\rho^{a(n)}+
    a^n_{mk}\left(\varepsilon^{abc}
    e_{\mu b}^{(m)}\tau_c^{(k)}+\varepsilon^{abc}\omega_{\mu b}^{(m)}
    \rho_c^{(k)}\right),
 \eea
where $a^n_{mk}$ defines the algebra structure with respect to 
the basis $\{e_m\}$.
If one defines the transformation parameter to be
 \bea\label{kkparam2}
  \rho^{a(n)}=a^n_{mk}\xi^{\mu(m)}e_{\mu}^{a(k)}, \qquad
  \tau^{a(n)}=a^n_{mk}\xi^{\mu(m)}\omega_{\mu}^{a(k)},
 \eea
one can show using the equations of motion and the associativity
(\ref{asso}) of the algebra, that
 \bea
  \delta e_{\mu}^{a(n)}=a_{mk}^n\left(\xi^{\rho(m)}\partial_{\rho}
  e_{\mu}^{a(k)}
  +\partial_{\mu}\xi^{\rho (m)}e_{\rho}^{a(k)}\right). 
 \eea
For the algebra-valued metric defined by 
$g_{\mu\nu}^n=a_{mk}^ne_{\mu}^{a(m)}e_{\nu a}^{(k)}$
this implies 
 \begin{equation}
  \begin{split}
   \delta_{\xi}g_{\mu\nu}^n&=\partial_{\mu}\xi^{\rho (l)}
   g_{\rho\nu\hspace{0.3em}l}^{\hspace{0.6em}n}+\partial_{\nu}
   \xi^{\rho (l)} g_{\rho\mu\hspace{0.3em}l}^{\hspace{0.6em}n}
   +\xi^{\rho (l)}\partial_{\rho}
   g_{\mu\nu\hspace{0.3em}l}^{\hspace{0.6em}n} \\
   &=\nabla_{\mu}\xi_{\nu}^n+\nabla_{\nu}\xi_{\mu}^n\;,
  \end{split}
 \end{equation}
where again (\ref{asso}) has been used. 
Altogether, the gauge transformations of the Chern-Simons theory
for (\ref{genalg}) coincide with the algebra-diffeomorphisms.
Thus we have shown that also for arbitrary internal manifolds 
the Chern-Simons description based on the algebra (\ref{genalg})
is equivalent to Wald's algebra-valued multi-graviton theory.

\end{appendix}

\end{document}